\documentclass[12pt,preprint]{aastex}
\usepackage[utf8]{inputenc}
\usepackage{graphicx}
\usepackage{natbib}
\usepackage{comment}
\usepackage{caption}

\newcommand{\kms}{\hbox{km~s$^{-1}$}}

\begin{document}

\title{Dual AGN candidates with double-peaked [O III] lines matching that of confirmed dual AGNs}

\author{
D.-C. Kim\altaffilmark{1}, Ilsang Yoon\altaffilmark{1}, A. S. Evans\altaffilmark{1,2}, Minjin Kim\altaffilmark{3}, E. Momjian\altaffilmark{4}, and Ji Hoon, Kim\altaffilmark{5} 
}
\altaffiltext{1}{National Radio Astronomy Observatory, 520 Edgemont Road, Charlottesville, VA 22903: dkim@nrao.edu, iyoon@nrao.edu, aevens@nrao.edu}
\altaffiltext{2}{Department of Astronomy, University of Virginia, 530 McCormick Road, Charlottesville, VA 22904}
\altaffiltext{3}{Department of Astronomy and Atmospheric Sciences, Kyungpook National University, Daegu 702-701, Republic of Korea, mkim.astro@gmail.com}
\altaffiltext{4}{National Radio Astronomy Observatory, P.O. Box O, Socorro, NM 87801, USA}
\altaffiltext{5}{Metaspace, 36 Nonhyeon-ro, Gangnam-gu, Seoul 06321, Republic of Korea; Subaru Telescope, National Astronomical Observatory of Japan, 650 North Aohoku Place, Hilo, HI 96720, USA, jhkim.astrosnu@gmail.com}

\begin{abstract}
We have performed a spectral decomposition to search for dual active galactic nuclei (DAGNs) in the Sloan Digital Sky Survey (SDSS) quasars with $z<0.25$.
Potential DAGN candidates are searched by referencing velocity offsets and spectral shapes of double-peaked [O III] lines of known DAGNs.
Out of 1271 SDSS quasars, we have identified 77 DAGN candidates.
Optical and mid-infrared diagnostic diagrams are used to investigate the ionizing source in the DAGN candidates.
The optical diagnostic analysis suggests 93\% of them are powered by AGNs, and mid-infrared diagnostic analysis suggests 97\% are powered by AGNs.
About 1/3 of the SDSS images of the DAGN candidates show signs of tidal interaction,
but we are unable to identify double nuclei in most of them due to the low spatial resolution of the archival imaging data available for most of the sample.
The radio-loud fraction of the DAGN candidates ($\sim$10\%) is similar to that of typical AGNs.

\end{abstract}


\section{Introduction}

In the current $\Lambda$ cold dark matter cosmology, 
galaxy interaction plays an important role in the evolution of galaxies:
it enhances strong starburst activity (Larson \& Tinsley 1978; Joseph et al. 1984; Sanders et al. 1988),
induces nuclear and AGN feedback (Lehnert \& Heckman 1996; Heckman et al. 2000; Rupke et al. 2002),
enriches the intergalactic medium with outflows (Nath \& Trentham 1997; Scannapieco et al. 2002),
and grows the mass of stellar bulges and supermassive black holes (SMBHs) (Kormendy \& Richstone 1995; Magorrian et al. 1998; Gebhardt et al. 2000; Ferrarese \& Merritt 2000; Tremaine et al. 2002; Marconi \& Hunt 2003; Hopkins et al. 2005).
Galaxy interaction also drastically alters internal and external structure of the host galaxies.
When two gas-rich galaxies start to approach each other, it induces long tidal tails in the outer parts of the galaxies
and develops bars in the inner parts of the galaxies.
The length of the tidal tails can reach up to a hundred kpc (i.e. Arp 188, Arp 295).
As the interaction continues, gas in the barred disks flows toward the central regions
and fuels kiloparsec-scale starbursts (Hernquist 1989; Barnes \& Hernquist 1996; De Rosa et al. 2018)
as seen in many luminous/ultraluminous infrared galaxies (LIRGs/ULIRGs).
Galaxy interaction is also thought to trigger star formation and fuel the central black holes and turn them into active galactic nuclei (AGNs).

Simulations suggest that merger-triggered dual AGNs 
can preferentially be found at the late phases of the mergers at the 1-10 kpc range of nuclear separation (Van Wassenhove et al. 2012).
While DAGNs in mergers are expected to be common, observationally they are rare and only $\sim$30 confirmed DAGNs have been found (Satyapal et al. 2017)
through optical (Liu et al. 2010, 2018; Woo et al. 2014; Huang et al. 2014; Bothun et al. 1989), mid-infrared (e.g., Tsai et al. 2013;
Ellison et al. 2017; Satyapal et al. 2017),
radio (Rodriguez et al. 2006; Frey et al. 2012; Owens et al. 1985), X-rays (Bianchi et al. 2008; Koss et al. 2011), 
optical + radio (M{\"u}ller-S{\'a}nchez et al. 2015; Fu et al. 2015), optical + X-rays (Comerford et al. 2011, 2015; Mazzarella et al. 2012; Ellison et al. 2017; Teng et al. 2012), and radio + X-rays (Komossa et al. 2003) observations.
The reason for finding a small number of DAGNs are not well known, but could be related to the AGN duty cycle, increased dust obscuration in the host galaxy center, or failures to detect them due to the bias of detection method.
The fate of DAGNs depends on a number of factors. They could merge together to form
a single supermassive black hole (SMBH), or could remain a dual black hole system if the
secondary galaxy loses significant fraction of its mass due to tidal stripping of
the primary galaxy at kpc scale (e.g. Callegari et al. 2009).
If two black holes in DAGNs are merging together, 
the merged SMBH can be recoiled and generate strong gravitational waves (Thorne \& Braginskii 1976; Campanelli et al. 2007; Schnittman 2007; Baker et al. 2008; Lousto \& Zlochower 2011).
Detailed studies of DAGNs and
their host galaxies can address the poorly constrained initial conditions for sub-pc binary SMBHs (Liu et al. 2018).

There have been systematic surveys for finding DAGN candidates, but most of the surveys are biased toward Type 2 AGNs (i.e. Wang et al. 2009; Liu et al. 2010; Ge et al. 2012).
In this paper, we report the results of searching Type 1 DAGN candidates based on the spectroscopic properties of known DAGNs.
In Section 2, the sample selection process for the DAGN candidates is described. 
The results and discussion are presented in Section 3,
and a summary of the paper is presented in Section 4.

\section{Sample selection process for DAGN candidates}

\subsection{Sample for DAGN candidates}
To search for DAGN candidates, we used a complete sample of 1271 $z < 0.25$ Sloan Digital Sky Survey (SDSS) SDSS Data Release 7 (DR7) quasar (Schneider et al. 2010).
The quasar activity is believed to be initiated by tidal interaction (i.e. Toomre \& Toomre 1972; Stockton 1990).
From our selection criteria, a quasar is already active in each candidate, and if the companion that triggered the quasar activity is also active, they will become either DAGNs or binary blackholes depending on their nuclear separation.
Therefore, using the quasar catalog to search for DAGNs could be a very efficient way to find them.
We will use a double-peaked [O III] $\lambda5007$ narrow emission line as a primary indicator of the DAGN candidates. 
The double-peaked [O III] narrow emission line can be produced by two distinct narrow-line regions (NLRs) in a dual or binary black hole systems.
It can also be produced by biconical outflows in a single black hole (Veilleux, Shopbell, \& Miller 2001; Liu et al. 2018).
The speed of AGN outflows will be decelerated with distance from the central AGN and
will produce a stratified velocity structure in narrow emission lines (i.e. lines with higher ionization potentials will be more blueshifted than that with lower ionization potential: Zamanov et al. 2002).
It is also possible that the double-peaked [O III] line can be produced by a rotating disk of NLR associated with a single black hole.
In this case, the NLR
gas is predominantly organized into a disk and produces the double-peaked [O III] line with comparable flux and line width (Greene \& Ho 2005).

\subsection{Spectral decomposition of known DAGNs and DAGN candidates}

A double-peaked [O III] line is easy to identify if the components are well separated as in gravitationally bound close binary black hole systems,
but if they are not gravitationally bound or are loosely bound, [O III] emission coming from each NLR could be blended together and thus difficult to identify in the spectra.
To increase detectability, we need to know how the spectral shapes of double-peaked [O III] line will look when two AGNs are well separated (i.e. the double-peaked [O III] line is blended). This task can be achieved if we artificially blend a well-separated double-peaked [O III] line in known DAGNs.
Satyapal et al. (2017) have compiled known DAGNs in their Table 8 and 
we have selected 11 DAGNs whose nuclear separation is
less than the SDSS fiber size of 3\arcsec \ so that both spectra from the DAGNs are detected in a single SDSS fiber aperture.
In addition, we have included a recently identified DAGN SDSS J092455.24+051052.0 that has a nuclear separation of 0.43" (Liu et al 2018).
Basic properties of these 12 known DAGNs are listed in Table 1.

Although it is expected that these 12 known DAGNs could have two [O III] $\lambda5007$\AA \ narrow line components,
we first started the spectral decomposition with one [O III] narrow line component.
We then performed the spectral decomposition with two [O III] narrow line components.
The results of the spectral decompositions with one [O III] component and two [O III] components are presented in the first and second panels in Fig. 1, respectively,
where the black, red, blue, and brown colors represent data, model, [O III] narrow line component, and power-law continuum, respectively. 
Their reduced $\chi^2$ values are listed in column 5 and column 6, respectively in Table 1.
Visual inspection of the resulting spectral decompositions, as well as $\chi^2$ values, suggest that,
except for J115822.58+323102.2, the spectral decomposition
with two [O III] components fits significantly better than that with one [O III] component.
The reduced $\chi^2$ values for the two [O III] component fit are at least 2.6$\times$ (J171544.05+600835.7) to up to 25.5$\times$ (JJ112659.54+294442.8) smaller than those for the one [O III] component fit.

Blue asymmetric profiles in the [O III] lines are often observed in AGNs (Heckman et al. 1981; Vrtilek \& Carleton 1985).
These asymmetric profiles are composed of a narrow line component and a blueshifted broad-line component (i.e. a wing component with FWHM of $\sim$500 \kms$-$1500 \kms).
The blue-shifted [O III] broad-line component is thought to originate from outflows and strong winds near the AGN 
(Heckman et al. 1981; Whittle et al. 1988; V{\' e}ron-Cetty et al. 2001; Bian et al. 2005; Komossa et al. 2008; Crenshaw et al. 2010; Cracco et al. 2016; Schmidt et al. 2018).
We performed spectral decomposition again with this broad-line component (green color), 
present the results in the third panels in Fig. 1, and list their reduced $\chi^2$ values in column 7 in Table 1.
Introducing the broad-line component yields much better fits than that of the two component fits
and decreases the reduced $\chi^2$ values from 1.3$\times$ (J112659.54+294442.8) to 13.9$\times$ (J095207.62+255257.2).
It also improves the fit significantly for J115822.58+323102.2; it decreases the reduced $\chi^2$ value from 9.28 (two component fit) to 0.73 (two component plus broad-line component fit).
Thus, we will use the results of spectral decomposition fitted with two [O III] narrow components plus blue-shifted broad-line component in the subsequent analysis.

The measured velocity offsets between the blue and red [O III] narrow emission lines in the SDSS spectra for known DAGNs 
range from 216 \kms \ to 632 \kms \ with a mean of 408$\pm$111 \kms.
Except for J112659.54+294442.8, we can visually identify double-peaked [O III] lines in the spectra. 
The instrumental resolution of SDSS spectra is 69 \kms.
The J112659.54+294442.8 has the smallest velocity offset (216 \kms $\sim$ 3$\times$ instrumental resolution) and 
spectral fitting with a single [O III] narrow emission line 
reduced $\chi^2$=1.16) is equally good as the fit with two [O III] narrow emission components (reduced $\chi^2$=0.91).
Thus, it is impossible to tell if this source is a DAGN or a single AGN without other details.
The next smallest velocity offset is for J114642.47+511029.6 (287 \kms $\sim$ 4$\times$ instrumental resolution).
In this case, we can visually identify the double-peaked [O III] lines.
It is expected that the distinguishability of double-peaked lines will depend on the FWHM of each line
(i.e. it will be easy to distinguish double-peaked lines if their FWHMs are small and will be hard if their FWHMs are large).
The FWHMs of J114642.47+511029.6 are 266 \kms \ and 276 \kms \ for the blueshift and redshift component, respectively.
Their FWHM values are somewhat smaller than the average value of FWHM of the total sample, which is 317 \kms.
Thus, we might $reliably$ identify double-peaked [O III] lines if their velocity offsets are larger than $\sim$4$\times$ instrumental resolution.
To test this, we have artificially translated the velocity offset between the blueshifted and redshifted [O III] lines to be
5$\times$ instrumental resolution (3rd panel in Fig. 1) and 4$\times$ instrumental resolution (4th panel in Fig. 1) from their original values.
In the 3rd panel (velocity offset = 5$\times$69 \kms), we can visually identify double peaks in 9 out of 12 sources.
For the remaining 3 sources (J110851.04+065901.4, J132323.33-015941.9, \& J150243.10+111557.3), we notice asymmetric Gaussian profiles in the spectral shapes that can be fitted with two Gaussian profiles.
If the velocity offset is 4$\times$69 \kms \ (4th panel in Fig. 1), we can still visually identify double-peaked lines for 5 out of 12 sources 
(J112659.54+294442.8, J114642.47+511029.6, J115822.58+323102.2, J142507.32+323137.4, \& J171544.05+600835.7; FWHM of them are narrower than that of the average value and their peak fluxes are comparable),
but it is difficult to tell if the spectral shape is composed of a single or double Gaussian lines for 4 of them (J102325.57+324348.4, J110851.04+065901.4, J132323.33-015941.9, \& J150243.10+111557.3).
For the remaining 3 sources (J092455.24+051052.0, J095207.62+255257.2, \& J162345.20+080851.1), 
we see a shoulder in a Gaussian profile due to one of the double-peaked components that can be revealed from spectral decomposition.
Therefore, the double-peaked [O III] line can be distinguished either from visual identification or spectral decomposition if the velocity offset is larger than 5$\times$69 \kms, 
and it can be identified for 2/3 of the spectra if velocity offset is about 4$\times$69 \kms.

The key parameters resulting from the spectral decompositions in Fig. 1 are 
i) the velocity offsets between blueshifted and redshifted [O III] components,
ii) the flux density ratios between blueshifted and redshifted [O III] components, and
iii) the strength of [O III] broad-line components.
After inspecting these parameters, we categorize spectral shapes of the known DAGNs into four classes (Fig. 2):
a) two [O III] narrow components can visually be identifiable and their flux density ratios are either larger (left plot) or smaller (right plot) than 1/2, and have a strong broad-line component,
b) same as a), except for having a weak broad-line component instead of a strong broad-line component,
c) a hint of a double-peaked [O III] line (either a shoulder in Gaussian profile or an asymmetric Gaussian profile) and their flux density ratios are either larger (left plot) or smaller (right plot) than 1/2, and have a strong broad-line component,
d) same as c), except for having a weak broad-line component.
We will use spectral shapes plotted in Fig. 2 as well as velocity offsets as references in the selection of DAGN candidates in the DR7 quasar spectra.

To select DAGN candidates, we have performed spectral decomposition on all SDSS DR7 quasars at $z<$0.25 
and have selected final candidates by referencing the above two selection criteria (velocity offset and spectral shape).
The spectral decomposition was performed with one [O III] plus broad-line components and with two [O III] plus broad-line components.
The results of the spectral decomposition with one [O III] plus broad-line components and two [O III] plus braod-line components 
are plotted in the left and right panels of Fig. 3 for each object, respectively, and their reduced $\chi^2$ values are listed in 4th and 5th columns in Table 2.
In all cases, as expected, the spectral decomposition with two [O III] components fits better than that with one [O III] component,
with the reduced $\chi^2$ values decreased from 1.4$\times$ (J114247.79+215709.5) to 26.4$\times$ (J122311.46+474426.7). 
In our final sample, we have included 3 candidates (J122313.21+540906.5, J123509.28+294849.9, \& J212203.82+001119.2) whose velocity offsets are between 200 \kms \ and 250 \kms 
\ since the results of their spectral decompositions were significantly better with two [O III] narrow line components than that with a single [O III] line component: 
the reduced $\chi^2$ values for J122313.21+540906.5, J123509.28+294849.9, \& J212203.82+001119.2 with the one [O III] plus broad-line components fit 
decrease from 12.32 to 0.59, 1.64 to 0.32, and 4.01 to 0.43, respectively, if we adopt the two [OIII] plus broad-line components fit.
In the right panel for each object, we have noted its matching spectral shape referenced to Fig. 2.
Fig. 4 shows the histogram of spectral shapes of the 77 DAGN candidates, where 22 (29\%) of them have visually identifiable double-peaked [O III] lines and 55 (71\%) have a hint of double-peaked [O III] lines, with their existence becoming more apparent after spectral decomposition.
The most common spectral shape of the DAGN candidates is class c) (41\%) and the next most common class is d) (32\%).

Histograms of velocity offsets and line widths of the known DAGNs (red color) and DAGN candidates (blue color) are plotted in Fig. 5.
The velocity offsets range from 216 \kms \ to 632 \kms\ for known DAGNs and 218 \kms \ to 854 \kms \ for DAGN candidates
with mean (median) velocity offsets of 406$\pm$118 \kms \ (408 \kms) for known DAGNs and 358$\pm$118 \kms \ (318 \kms) for DAGN candidates.
The FWHMs of [O III] lines range from 175 \kms \ to 472 \kms\ for known DAGNs and 220 \kms \ to 747 \kms \ for DAGN candidates,
with mean (median) FWHMs of 317$\pm$67 \kms (312 \kms) for known DAGNs and 407$\pm$94 \kms \ (396 \kms) for DAGN candidates.
If we consider the standard deviation of the mean velocity offset and the FWHM, the known dAGN and dAGN candidates have the similar velocity offsets and FWHM of [OIII] line.



\section{Discussion}

We have identified DAGN candidates based on the spectral properties of the known DAGNs.
Here, we can ask `does the double-peaked [O III] line in DAGN candidates really originate from dual AGNs?'.
To answer this question, 
we investigate optical and mid-infrared diagnostic line diagrams to identify ionization sources (i.e. star forming galaxies (SFG) vs. AGN), and
we examine SDSS, Hubble Space Telescope (HST), Very Large Array Sky Survey (VLASS) images to identify double nuclei in the host galaxy. 

\subsection{Diagnostic line diagrams}
\subsubsection{Optical diagnostic line diagram}

In the optical range, the BPT-VO diagnostic line diagrams (Baldwin, Phillips, \& Terlevich 1981; Veilleux \& Osterbrock 1987; Kewley et al. 2006) 
are widely used to distinguish ionization mechanisms.
We will use line ratios of [O III]/H$\beta$ and [N II]/H$\alpha$ to distinguish AGNs and SFGs (i.e. H II region galaxies).
First, we will examine line ratios of the known DAGNs.
Integrated galaxy line ratios for the known DAGNs are plotted in Fig. 6a,
where numbers in parenthesis represent the velocity offsets between the blueshifted and redshifted [O III] components and 
the projected nuclear separation between the two nuclei, respectively.
The black circles with magenta color represent radio sources detected by Very Large Array Sky Survey (VLASS, Lacy et al. 2020). 
As expected, we find that all sources are powered by AGNs.
We also find that the strengh of AGN activity (more AGN-like or less AGN-like) does not have any dependencies on the velocity offset, nuclear separation, or radio activity.
The line ratios of individual nuclei are plotted in Fig. 6b, where blue and red circles represent line ratios for [O III] blueshifted component and [O III] redshifted component, respectively, and a line is connecting between the two line ratios.
The plot tells us that both the blueshifted and redshifted [O III] components have line ratios consistent with AGN.
There is suggestive evidence that the redshifted [O III] component is more AGN-like (positive [O III]/H$\beta$ direction) than the blueshifted [O III] component.

Integrated galaxy line ratios for DAGN candidates are plotted in Fig. 7a.
None of the sources fall in the SFG category: 93\% (71 out of 77) falls in the AGN category and 7\% (5 out of 77) fall in the composite (i.e. AGN+SFG) category.
This result is expected since we used SDSS quasar data to search for DAGN candidates.
In Figure 7b, we examine whether the line ratios of individual nuclei are also powered by AGNs. 
For 20 DAGN candidates (26\%), we were able to measure H$\beta$, H$\alpha$, and [N II] fluxes for each blueshifted and redshifted [O III] component.
Again, none of the sources fall in the H II region galaxy category; 93\% (37 out of 40) fall in the AGN category and 7\% (3 out of 40) fall in the composite category.

\subsubsection{Mid infrared diagnostic line diagram}
The mid-infrared diagnostic line ratios can be used to distinguish SFGs and AGNs even in cases 
where one or both of the AGNs are optically obscured (Lacy et al. 2004; Stern et al. 2005; Coziol, Torres-Papaqui, \& Andernach 2015).
The Wide-field Infrared Survey Explorer (WISE; Wright et al. 2010) performed
an all-sky photometric survey in four wave bands in the mid-infrared (MIR): 3.4 $\mu$m (W1), 4.6 $\mu$m (W2), 12 $\mu$m (W3), and 22 $\mu$m (W4). 
For the mid-infrared diagnostic line diagram, we will use W2-W3 and W3-W4 colors.
The mid-infrared diagnostic line diagram for known DAGNs is plotted in Fig. 8a, where the solid line divides AGN and SFG, and 
brown, magenta, blue, green, orange, and cyan rectangles represent the first and third quartile boundaries of box and whisker plot 
of (W2-W3) and (W3-W4) colors (Coziol, Torres-Papaqui, \& Andernach 2015) for quasar, Sey 1, Sey 2, LINER, Composite, and SFG, respectively.
Two DAGNs are on the border line of the AGN and SFG while the rest of the known DAGNs fall into the AGN category.
Interestingly, the most AGN-like source (J112659.54+294442.8: upper-left data point) in this plot corresponds to the most AGN-like source in BPT-VO diagram (top data point), and
the least AGN-like source (J092455.24+051052.0: lowest data point) in this plot corresponds to the least AGN-like source in BPT-VO plot (bottom data point).
The mid-infrared diagnostic line diagram for DAGN candidates is plotted in Fig. 8b.  
All but 2 fall AGN category, where more than half of them fall quasar/Sey1/Sey2 sub-categories.
Overall, the mid-infared diagnostic line diagram for known DAGNs and DAGN candidates yields comparable results to that of the optical diagnostic line diagram.

\subsection{Imaging properties}

The optical and mid-infrared diagnostic diagrams suggest most of the double-peaked [O III] lines originate from AGNs.
If they are indeed DAGNs, we may detect double nuclei that are AGNs in the galaxy center.
Even if the two nuclei cannot be detected
due to dust obscuration, signs of tidal interactions could be indirect evidence that the
double-peaked [O III] lines could be powered by AGNs.
To test this, we investigate available images of the known DAGNs and DAGN candidates.

\subsubsection{Images of known DAGNs}
For known DAGNs, we expect to detect double nuclei in the center of each galaxy unless they are heavily obscured
since the double-peaked [O III] lines are already confirmed to be originated from the dual AGNs.
Fig. 9 shows SDSS images of the known DAGNs.
We find only one object has a double nucleus in its center (J114642.47+511029.6).
Signs of tidal interactions can be seen in J092455.24+051052.0, J110851.04+065901.4, and J150243.10+111557.3.
For the rest of the objects, it is impossible to tell the existence of double nuclei or any signs of tidal interactions due to the low spatial resolution of SDSS.
This situation can be improved if we have high resolution images, and for this we have searched the HST archive and found images of 7 out of 12 known DAGNs.
Fig. 10 shows the HST images, where the HST camera and filter used are provided on top of each plot just below the SDSS source name, and insets show blown-up images of the nuclear region.
Double nuclei can be seen in all but one image (J171544.05+600835.7).
Note that J092455.24+051052.0 (Liu et al. 2018) and J132323.33-015941.9 (Woo et al. 2014) are very close doubles,
and J112659.54+294442.8 has a point-like source $\sim$1\arcsec \ SE of the central nucleus which could be the remnant black hole from a minor merger interaction with the main galaxy (Comerford et al. 2015).
Dual AGNs separated by 1.9 kpc were detected in J171544.05+600835.7 with Chandra X-ray observations (Comerford et al. 2011),
but we do not see double nuclei in the HST image probably because of dust obscuration.
In the HST images, host morphologies of all but one source show signs of interactions/mergers that are not apparent in the low resolution SDSS images.

The existence of dual AGNs can be identified with radio observations in the form of two flat-spectrum compact cores,
multiple jets with misaligned axis, or X-shaped radio emissions.
Fig. 11 shows the VLASS images of the known DAGNs. The FOV is 12\arcsec\ and the lowest contour level is 3$\times$RMS noise (1 RMS=1.2$\times 10^{-4}$ Jy).
The resolution of the VLASS is 2\farcs5, hence it is not possible to identify the compact cores in each galaxy, but larger scale radio jets may be seen.
The VLASS images do not show apparent radio jets, except a faint jet-like structure stretching from 
center to the north \& south directions in J102325.57+324348.4. 
The radio-loud ($R$=$L_{5GHz}$/$L_B>10$: Kellermann et al. 1989) fraction of the known DAGNs was found to be significantly larger (50\%) than that of typical AGNs (15\% - Kellermann et al. 1989; Gupta, Sikora, \& Rusinek 2020).
For the confirmation of two AGNs in our known DAGN sample, radio observations 
(VLA: J102325.57+324348.4, J115822.58+323102.2, J150243.10+111557.3; VLBI: J142507.32+323137.4) were made use of for 4 out of 6 radio-loud AGNs.
This might explain a high fraction of radio-loud AGNs in our sample.
However, the known DAGNs used in this work as a reference for spectral shape are not from homogeneous selection criteria and therefore their radio-loudness does not have a statistical meaning.  

\subsubsection{Images of DAGN candidates}

The SDSS images of 77 DAGN candidates are plotted in Fig. 12.
About 1/3 of the sources show tidal tails and about 20\% have companions within the 12\arcsec \ FOV of which only 3 sources have double nuclei within $\sim$3\arcsec\ diameter (J080544.13+113040.2, J123557.86+582122.9, \& J125016.21+045745.0).
As in the SDSS images for known SDSS DAGNs, we could not identify dual AGNs for the rest of the sources due to low spatial resolution.
We have searched the HST archive and found images for only 4 DAGN candidates (Fig. 13).
J115349.26+112830.4 has very bright core and does not appear to have a double nuclei in the center.
It has two faint sources $\sim$3\farcs8 \ (14.4 kpc, $r$=21.1 mag) in the north-east and $\sim$7\farcs1\ (26.9 kpc, $r$=20.0 mag) in the south-west directions, 
but it is not clear if they are companions or foreground/background sources.
J123915.40+531414.6 has a double nuclei in the center and companion-like object in the north-east direction.
In the SDSS image of J154307.77+193751.7, we see a disturbed plume-like structure in the south-west direction.
This structure appears to be a thick tidal tail in the HST image.
We do not find any signs of tidal interaction in the SDSS image of J165939.77+183436.8,
but a faint disturbed envelope in the outskirt of the galaxy and a double nuclei in the center of the galaxy are found in the HST image.
The VLASS images of the DAGN candidates are shown in Fig. 14.
We see sources with double radio cores in J122313.21+540906.5 (possibly in J141041.50+223337.0 too),
very strong radio jets in J094822.45+684835.2, J130007.99+035556.6, \& J142040.86+065059.5, and
moderate-intensity jets in J100208.14+345353.7\& J103359.46+355509.0.
Unlike known DAGNs, the radio-loud fraction of the DAGN candidates is $\sim10$\%, which is similar to that of the typical AGNs.

\subsection{Comparison with other surveys}

We note that there are previous studies of AGNs with double peaked narrow emission lines. 
Wang et al. (2009), Liu et al. (2010), Smith et al. (2010), and Ge et al. (2012) performed DAGNs searches using the SDSS DR7 catalog.
However, Wang et al. (2009), Liu et al. (2010), and Ge et al. (2012) searched the DAGNs using SDSS DR7 narrow emission line galaxies, 
thus we will not compare their results with ours since we used quasars.
We confirmed that there are no overlapping DAGN candidates between their lists and ours.
Smith et al. (2010) have identified 148 DAGN candidates by means of a visual examination of 21,592 SDSS DR7 quasars at $0.1<z<0.7$.
We have compared our list with theirs and found
they have not identified 69 DAGN candidates we have identified and we have missed 1 DAGN candidate (J171930.56+293412.8) they have identified.
We have inspected the spectra of J171930.56+293412.8 and found that the spectral shape of the double-peaked [O III] line 
appeared to be a mixture of multiple Gaussians and does not look like a typical [O III] line as shown in Fig. 1 or Fig. 2.
Furthermore, the line width of the J171930.56+293412.8 is about 3 to 4 times (FWHM([O III] blueshifted component)$\sim$1,200\kms \ and FWHM([O III] redshifted component)$\sim$900\kms) 
broader than that of the known DAGNs and thus would not have been included in our list.
We suspect that they may have picked DAGN candidates that have large velocity offsets since they have identified the double-peaked [O III] line via visual inspection.
Our assumption seems to be partly correct since the mean velocity offset for the 7 overlapping DAGN candidates (467$\pm$203 \kms) is $\sim$120 \kms \ larger than that of the rest of 70 (346$\pm$101 \kms).

\section{Summary}

DAGN candidates have been identified via spectral decomposition of SDSS DR7 quasars with $z<$0.25.
To identify DAGN candidates, the spectral shapes and velocity offsets of double-peaked [O III] lines of known DAGNs are used as a reference.
We found 6\% (77 out of 1271 sources) of SDSS quasars have double-peaked [O III] lines and their spectral shapes are similar to those of known DAGNs.
The mean velocity offset and FWHM of the DAGN candidates are 358$\pm$118 \kms \ and 407$\pm$94 \kms, respectively.
Optical and mid-infrared diagnostic diagrams are used to investigate ionizing sources in the DAGN candidates.
BPT-VO diagnostic diagram suggests 93\% of them are powered by AGNs and mid-infrared diagnostic diagram suggests 97\% are powered by AGNs. 
Optical and radio images are used to find dual AGNs in galaxy centers.
About 1/3 of the SDSS images of the DAGN candidates show signs of tidal interaction,
but except for 3 sources we were unable to identify double nuclei because of poor spatial resolution.
The HST archive was searched for higher resolution images, and 4 HST images are found of which one of the DAGN candidates has a thick tidal tail and the other two have double nuclei.
For the known DAGNs, the radio-loud fraction is 50\% and for the candidate DAGNs, the radio-loud fraction is 10\%.
The large radio-loud fraction in the known DAGNs could be due to the fact that radio observations are used to confirm the existence of DAGNs. 
It could also be possible that a sizable fraction of DAGN candidates may not be genuine DAGNs 
if high radio loud fractions in known DAGNs is an intrinsic property.
However more importantly, a careful study of the radio properties of genuine DAGNs that are 'radio' unbiased is necessary in order to confirm the correlation (or anti-correlation) of radio-loudness with dual AGN activity. 
Even though optical and mid-infrared diagnostic diagrams suggest most of them are powered by AGNs,
high resolution images are strongly needed to identify double nuclei in these DAGN candidates to confirm if they are a real DAGNs.
The confirmed quasar DAGNs will be an important complementary to the narrow emission line DAGNs and will serve as a valuable sample for studying the AGN activities and black hole growth.

\section{Acknowledgement}
The authors thank the anonymous referee for his/her useful comments and suggestions.
This research has made use of the NASA/IPAC Extragalactic Database
(NED) which is operated by the Jet Propulsion Laboratory, California
Institute of Technology, under contract with the National Aeronautics and Space Administration. 
D.K., I.Y., A.E., and E.M. acknowledge support from the National Radio Astronomy Observatory (NRAO) and 
M.K. was supported by the National Research Foundation of Korea (NRF) grant funded by the Korea government (MSIT) (No. 2020R1A2C4001753).
The National Radio Astronomy Observatory is a facility of the National Science Foundation operated under cooperative agreement by Associated Universities, Inc.

\clearpage

\begin{deluxetable}{lccccccccccc}
\tabletypesize{\scriptsize}
\tablewidth{0pt}
\tablecaption{Properties of the Known DAGNs}
\tablehead{
\multicolumn{1}{c}{Name} &
\multicolumn{1}{c}{z} &
\multicolumn{1}{c}{r} &
\multicolumn{1}{c}{Nuc. Separation} &
\multicolumn{3}{c}{Reduced $\chi^2$} &
\multicolumn{1}{c}{$\Delta V$([O III])} &
\multicolumn{1}{c}{VLASS} &
\multicolumn{1}{c}{Radio Loud} \\
\cline{5-7}
\multicolumn{1}{c}{SDSS} &
\multicolumn{1}{c}{} &
\multicolumn{1}{c}{mag} &
\multicolumn{1}{c}{\arcsec\ \ (kpc)} &
\multicolumn{1}{c}{1c} &
\multicolumn{1}{c}{2c} &
\multicolumn{1}{c}{2c+bc} &
\multicolumn{1}{c}{\kms} &
\multicolumn{1}{c}{} &
\multicolumn{1}{c}{}\\
\multicolumn{1}{c}{(1)} &
\multicolumn{1}{c}{(2)} &
\multicolumn{1}{c}{(3)} &
\multicolumn{1}{c}{(4)} &
\multicolumn{1}{c}{(5)} &
\multicolumn{1}{c}{(6)} &
\multicolumn{1}{c}{(7)} &
\multicolumn{1}{c}{(8)} &
\multicolumn{1}{c}{(9)} &
\multicolumn{1}{c}{(10)}
}
\startdata
	$J092455.24+051052.0$ & 0.150 & 16.90 & 0.45 (1.1) &2.63  & 0.32 & 0.21 & 420 & N & N \\
	$J095207.62+255257.2$ & 0.339 & 18.22 & 1.00 (4.8) &9.66  & 3.19 & 0.23 & 434 & N & N \\
	$J102325.57+324348.4$ & 0.127 & 16.95 & 0.45 (1.0) &3.98  &0.89 & 0.52 & 383 & Y & Y \\
	$J110851.04+065901.4$ & 0.182 & 17.08 & 0.69 (2.1) &28.83 &4.42 & 1.96 & 396 & Y & Y \\
	$J112659.54+294442.8$ & 0.102 & 16.84 & 1.17 (2.2) &29.59 &1.16 & 0.91 & 317 & N & N \\
	$J114642.47+511029.6$ & 0.130 & 17.42 & 2.72 (6.3) &13.42 &2.34 & 1.14 & 287 & N & N \\
	$J115822.58+323102.2$ & 0.166 & 17.06 & 0.22 (0.6) &4.64  &9.28 & 0.73 & 448 & Y & Y \\
	$J132323.33-015941.9$ & 0.350 & 19.59 & 0.16 (0.8) &7.66  &1.21 & 0.46 & 216 & Y & N \\
	$J142507.32+323137.4$ & 0.478 & 18.22 & 0.44 (2.6) &12.22 &1.65 & 0.34 & 491 & Y & Y \\
	$J150243.10+111557.3$ & 0.391 & 18.02 & 1.40 (7.4) &58.39 &10.37 & 1.66 & 632 & Y & Y \\
	$J162345.20+080851.1$ & 0.199 & 17.28 & 0.47 (1.6) &12.80 &1.88 & 0.78 & 502 & Y & N \\
	$J171544.05+600835.7$ & 0.157 & 17.35 & 0.70 (1.9) &22.97 &8.80 & 1.84 & 349 & Y & Y \\
\enddata

\tablenotetext{\ } {{\it Col 1:}\ Object name.}
\tablenotetext{\ } {{\it Col 2:}\ Redshift.}
\tablenotetext{\ } {{\it Col 3:}\ SDSS $r$-band magnitude.}
\tablenotetext{\ } {{\it Col 4:}\ Projected nuclear separation in arcsec (kpc) units.}
\tablenotetext{\ } {{\it Col 5$-$7:}\ Reduced $\chi^2$ value with one [O III] component fit, two [O III] components fit, and two [O III] components plus [O III] broad-line component fit, respectively.}
\tablenotetext{\ } {{\it Col 6:}\ Projected nuclear separation in arcsec (kpc) units.}
\tablenotetext{\ } {{\it Col 7:}\ Projected nuclear separation in arcsec (kpc) units.}
\tablenotetext{\ } {{\it Col 8:}\ Velocity offset between the redshifted and blueshifted [O III] lines.}
\tablenotetext{\ } {{\it Col 9:}\ Radio emission detected from VLASS.}
\tablenotetext{\ } {{\it Col 10:}\ Radio loudness $R$ estimated from 5 GHz luminosity to blue luminosity ratio (Radio-loud: $R>10$).}

\end{deluxetable}

\begin{deluxetable}{lccccc|cccccc}
\tabletypesize{\tiny}
\tablewidth{0pt}
\tablecaption{Properties of the DAGN Candidates}
\tablehead{
\multicolumn{1}{c}{Name} &
\multicolumn{1}{c}{z} &
\multicolumn{1}{c}{r} &
\multicolumn{2}{c}{Reduced $\chi^2$} &
\multicolumn{1}{c|}{$\Delta V$} &
\multicolumn{1}{c}{Name} &
\multicolumn{1}{c}{z} &
\multicolumn{1}{c}{r} &
\multicolumn{2}{c}{Reduced $\chi^2$} &
\multicolumn{1}{c}{$\Delta V$} \\
	\cline{4-5}
	\cline{10-11}
\multicolumn{1}{c}{SDSS} &
\multicolumn{1}{c}{} &
\multicolumn{1}{c}{mag} &
\multicolumn{1}{c}{1c+bc} &
\multicolumn{1}{c}{2c+bc} &
\multicolumn{1}{c|}{\kms} &
\multicolumn{1}{c}{SDSS} &
\multicolumn{1}{c}{} &
\multicolumn{1}{c}{mag} &
\multicolumn{1}{c}{1c+bc} &
\multicolumn{1}{c}{2c+bc} &
\multicolumn{1}{c}{\kms} \\
\multicolumn{1}{c}{(1)} &
\multicolumn{1}{c}{(2)} &
\multicolumn{1}{c}{(3)} &
\multicolumn{1}{c}{(4)} &
\multicolumn{1}{c}{(5)} &
\multicolumn{1}{c|}{(6)} &
\multicolumn{1}{c}{(1)} &
\multicolumn{1}{c}{(2)} &
\multicolumn{1}{c}{(3)} &
\multicolumn{1}{c}{(4)} &
\multicolumn{1}{c}{(5)} &
\multicolumn{1}{c}{(6)}
}
\startdata
$J080544.13+113040.2$ & 0.1991 & 17.43 & 10.25&0.69&455 & $J131256.75+163150.5$ & 0.1646 & 17.31 & 3.76& 1.30&252 \\
$J081359.30+213512.7$ & 0.1800 & 16.80 & 0.33& 0.23&380 & $J132105.98+504634.4$ & 0.2330 & 17.35 & 0.14& 0.10&256 \\
$J081542.53+063522.9$ & 0.2438 & 17.72 & 0.62& 0.14&338 & $J132508.60+061112.2$ & 0.1831 & 17.95 & 3.19& 0.75&256 \\
$J085431.28-003650.6$ & 0.2370 & 18.08 & 0.88& 0.20&269 & $J132832.58-023321.4$ & 0.1836 & 16.85 & 1.18& 0.13&313 \\
$J085632.40+504114.0$ & 0.2346 & 16.32 & 1.05& 0.50&724 & $J132931.37+202723.1$ & 0.2314 & 18.05 & 0.62& 0.18&255 \\
$J090427.29+374357.4$ & 0.1983 & 17.48 & 0.18& 0.09&271 & $J134327.27+082234.1$ & 0.2426 & 17.51 & 1.15& 0.16&378 \\
$J092140.83+293809.6$ & 0.2262 & 17.45 & 2.00& 0.53&272 & $J134615.88+580008.1$ & 0.1625 & 16.90 & 2.05& 0.70&403 \\
$J092635.11+072446.4$ & 0.1894 & 17.75 & 3.30& 0.26&453 & $J135550.20+204614.5$ & 0.1962 & 15.69 & 2.71& 0.18&566 \\
$J092820.98+493736.8$ & 0.2383 & 17.68 & 0.43 &0.15&270 & $J140208.86+462414.4$ & 0.2369 & 17.96 & 0.62& 0.25&287 \\
$J094439.88+034940.1$ & 0.1554 & 16.62 & 0.39& 0.10&318 & $J140914.35+565625.7$ & 0.2386 & 17.91 & 0.21& 0.12&345 \\
$J094822.45+684835.2$ & 0.2020 & 17.86 & 13.32&1.37&278 & $J141041.50+223337.0$ & 0.1728 & 16.79 & 1.48& 0.54&328 \\
$J100208.14+345353.7$ & 0.2052 & 17.25 & 1.12& 0.21&388 & $J141053.43+091026.9$ & 0.1779 & 16.56 & 3.75& 0.21&314 \\
$J100627.94+603043.6$ & 0.2103 & 17.66 & 0.14& 0.08&345 & $J141315.28+290637.7$ & 0.2252 & 17.55 & 0.78& 0.37&406 \\
$J100642.57+412201.9$ & 0.1498 & 16.46 & 2.09& 0.66&316 & $J141556.84+052029.5$ & 0.1264 & 16.76 & 3.83& 0.31&258 \\
$J101312.15+020416.4$ & 0.2201 & 17.16 & 0.21& 0.08&292 & $J142040.86+065059.5$ & 0.2375 & 17.52 & 0.21& 0.07&253 \\
$J102152.34+464515.6$ & 0.2043 & 17.43 & 0.18& 0.09&257 & $J143422.97+324041.4$ & 0.2455 & 17.78 & 0.31& 0.13&317 \\
$J103359.46+355509.0$ & 0.1688 & 16.67 & 4.11& 0.22&494 & $J144347.48+513050.0$ & 0.2005 & 17.95 & 0.57& 0.28&306 \\
$J103808.63+233133.4$ & 0.2260 & 17.53 & 4.48& 1.51&301 & $J144748.79+624444.7$ & 0.2299 & 17.67 & 0.39& 0.21&639 \\
$J103953.89+293847.8$ & 0.2461 & 17.52 & 1.20& 0.60&277 & $J145434.34+080336.7$ & 0.1298 & 16.26 & 2.43& 0.12&350 \\
$J104111.97+282805.0$ & 0.2110 & 16.39 & 0.33& 0.11&277 & $J151907.33+520605.9$ & 0.1379 & 16.61 & 3.84& 0.55&267 \\
$J113020.99+022211.5$ & 0.2412 & 17.51 & 1.00& 0.22&400 & $J152422.14+264223.6$ & 0.2365 & 18.07 & 0.99& 0.09&322 \\
$J113257.84+604653.6$ & 0.2325 & 18.35 & 2.46& 0.61&305 & $J153231.80+420342.7$ & 0.2095 & 18.08 & 9.15& 0.37&354 \\
$J114247.79+215709.5$ & 0.2309 & 17.14 & 0.18& 0.13&315 & $J154307.77+193751.7$ & 0.2289 & 16.68 & 1.24& 0.11&482 \\
$J115349.26+112830.4$ & 0.1766 & 16.36 & 1.93& 0.19&359 & $J154316.41+540526.0$ & 0.2452 & 17.79 & 0.25& 0.14&253 \\
$J121522.77+414620.9$ & 0.1961 & 17.52 & 3.52& 0.16&389 & $J154518.05+463837.9$ & 0.2282 & 17.17 & 1.81& 0.26&424 \\
$J122311.46+474426.7$ & 0.1627 & 16.43 & 1.57& 0.28&336 & $J154732.17+102451.2$ & 0.1381 & 15.91 & 1.70& 0.28&488 \\
$J122313.21+540906.5$ & 0.1559 & 16.62 & 12.32&0.59&232 & $J155318.72+170202.9$ & 0.1611 & 17.19 & 0.21& 0.04&360 \\
$J123341.80+644317.4$ & 0.2221 & 16.81 & 0.53& 0.13&253 & $J155446.71+362141.4$ & 0.2494 & 18.35 & 0.50& 0.17&294 \\
$J123509.28+294849.9$ & 0.2430 & 17.81 & 1.64& 0.32&231 & $J160518.49+375653.4$ & 0.2009 & 17.04 & 0.13& 0.06&270 \\
$J123557.86+582122.9$ & 0.2115 & 17.94 & 6.25& 0.43&301 & $J161649.37-004250.4$ & 0.2379 & 16.83 & 3.70& 0.23&459 \\
$J123915.40+531414.6$ & 0.2013 & 16.71 & 1.66& 0.69&854 & $J162633.92+480230.1$ & 0.2426 & 17.47 & 0.34& 0.10&451 \\
$J124013.05+310243.3$ & 0.2357 & 17.71 & 0.27& 0.11&305 & $J164331.90+304835.5$ & 0.1837 & 17.11 & 0.43& 0.30&373 \\
$J124238.37+045616.5$ & 0.2338 & 18.14 & 1.51& 0.18&396 & $J165939.77+183436.8$ & 0.1708 & 17.11 & 5.03& 1.58&263 \\
$J124323.82+051446.3$ & 0.1642 & 16.64 & 0.28& 0.15&344 & $J171411.63+575833.9$ & 0.0926 & 15.88 & 0.50& 0.06&290 \\
$J124813.82+362423.6$ & 0.2069 & 17.40 & 7.93& 0.30&377 & $J171526.52+291923.5$ & 0.2082 & 18.00 & 0.20& 0.11&254 \\
$J125016.21+045745.0$ & 0.2337 & 17.61 & 0.46& 0.15&325 & $J212203.82+001119.2$ & 0.2289 & 18.61 & 4.01& 0.43&218 \\
$J125051.04+060909.9$ & 0.1821 & 16.97 & 2.25& 0.24&603 & $J222909.81+002527.3$ & 0.2276 & 17.84 & 0.16& 0.06&489 \\
$J130007.99+035556.6$ & 0.1841 & 17.15 & 2.95& 0.39&311 & $J230155.55-010649.0$ & 0.2383 & 16.95 & 1.11& 0.05&601 \\
$J130416.99+020537.0$ & 0.2286 & 17.18 & 0.19& 0.08&526 &                       &        &       &  &  &   \\
\enddata

\tablenotetext{\ } {{\it Col 1:}\ Object name.}
\tablenotetext{\ } {{\it Col 2:}\ Redshift.}
\tablenotetext{\ } {{\it Col 3:}\ SDSS $r$-band magnitude.}
\tablenotetext{\ } {{\it Col 4$-$5:}\ Reduced $\chi^2$ value with one [O III] component plus [O III] broad-line component fit, and two [O III] components plus [O III] broad-line component fit, respectively.}
\tablenotetext{\ } {{\it Col 6:}\ Velocity offset between the redshifted and blueshifted [O III] lines.}

\end{deluxetable}

\clearpage

\begin{figure}[!hb]
\centerline{\includegraphics[scale=0.9]{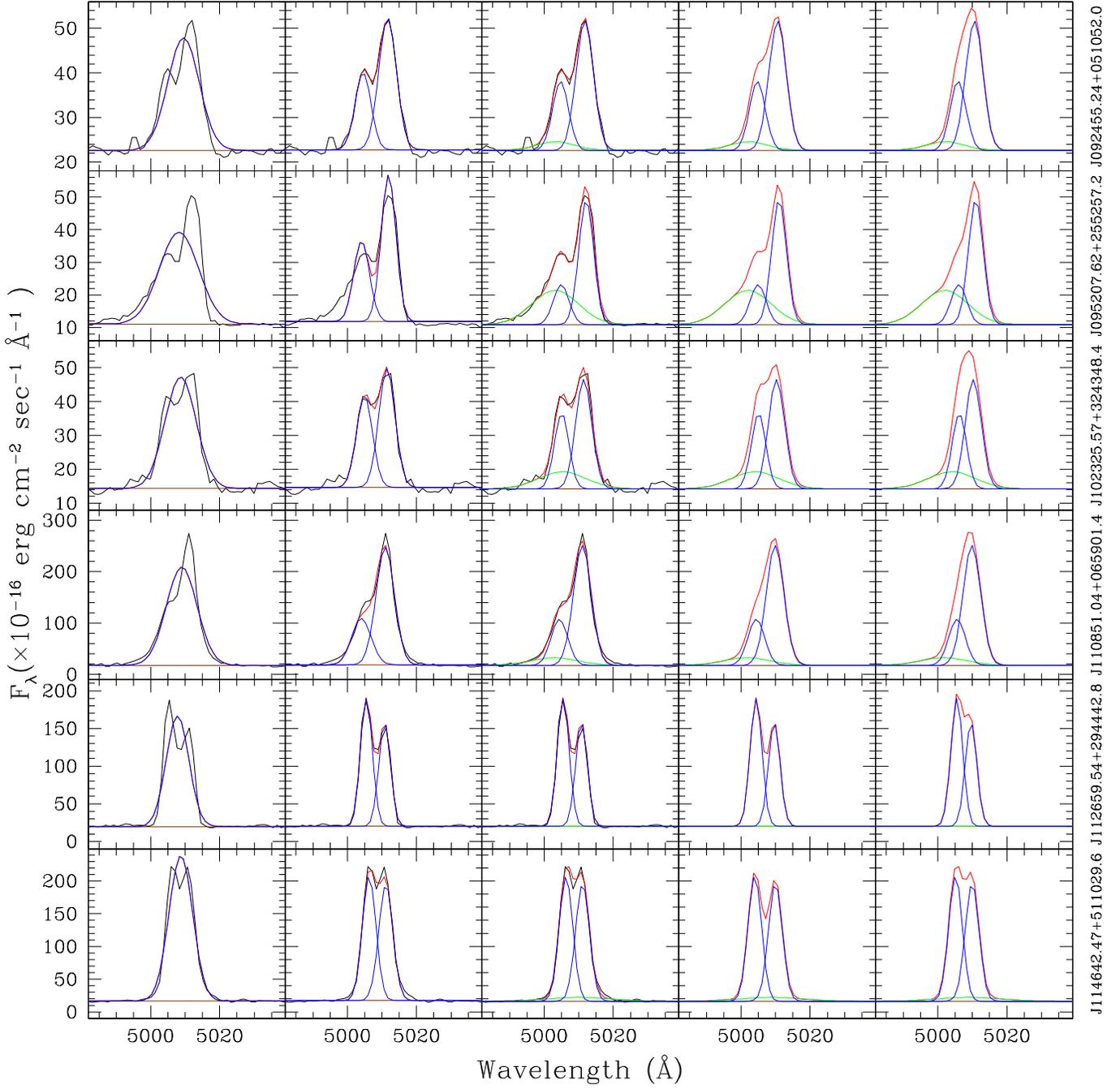}}
	\caption{Spectral decomposition of [O III]$\lambda 5007$ for known DAGNs, where black, red, brown, blue, and green represent data, model, continuum, narrow line component, and broad-line component, respectively. 
The 1st, 2nd, and 3rd panel shows spectral decomposition with one [O III] component, two [O III] components,
and two [O III] plus broad-line components, respectively. The 4th and 5th panel represent model spectra with the velocity offset of two [OIII] narrow emission lines by 5$\times$instrumental resolution and 4$\times$instrumental resolution, respectively.
}
\end{figure}

\begin{figure}[!hb]
\centerline{\includegraphics[scale=0.9]{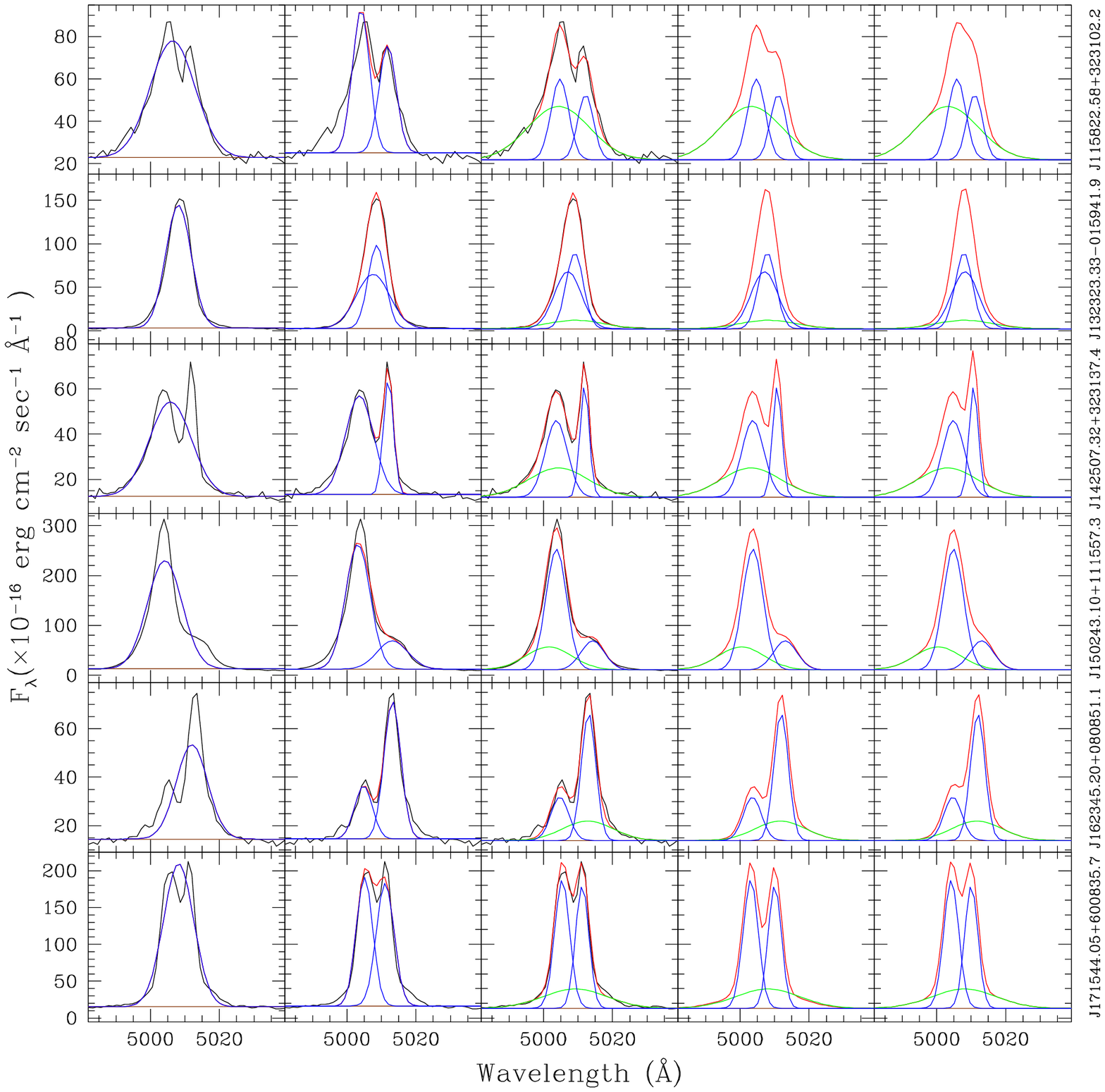}}
\end{figure}

\begin{figure}[!hb]
\centerline{\includegraphics[scale=1.0]{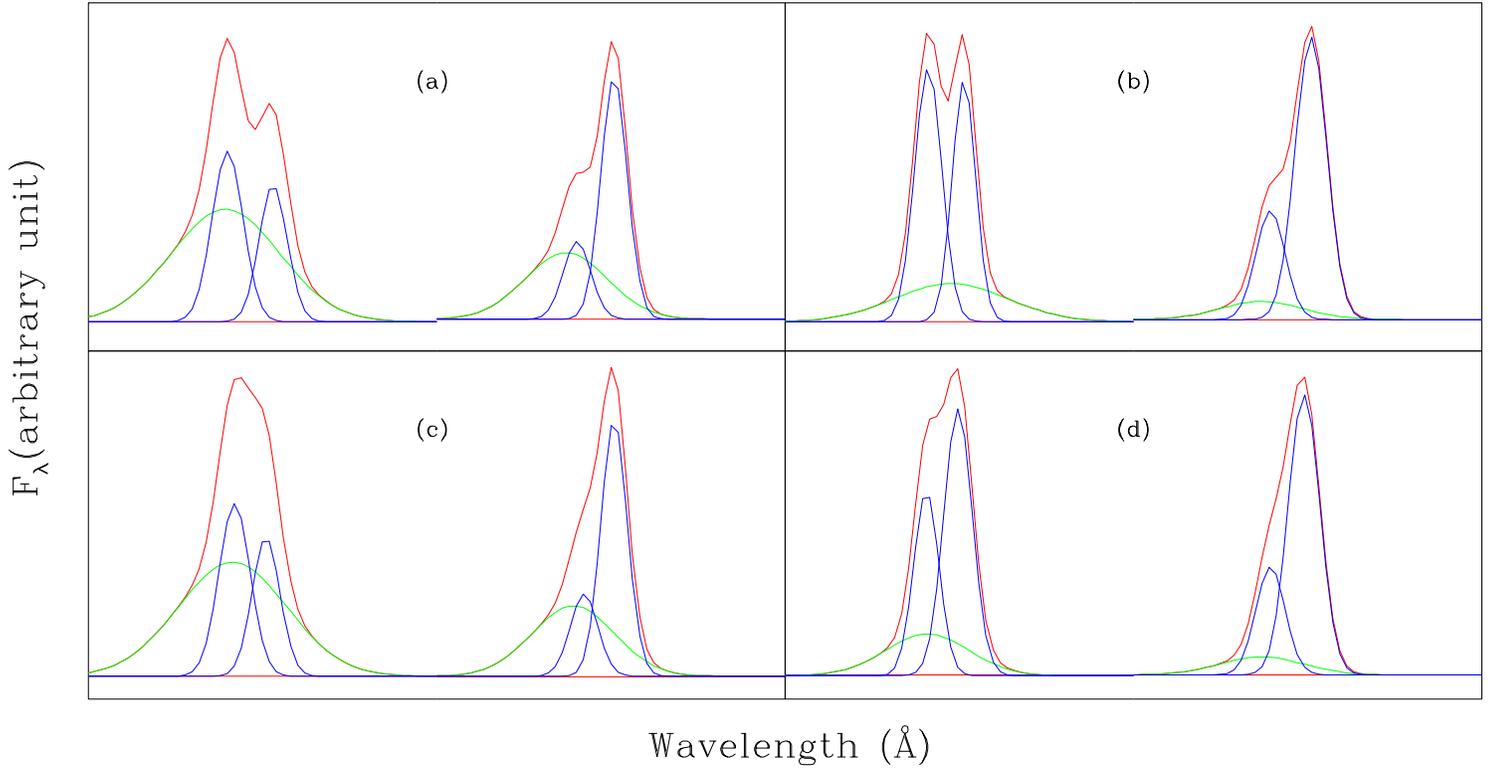}}
\caption{Representative spectral shape of known DAGNs
}
\end{figure}

\begin{figure}[!hb]
\centerline{\includegraphics[scale=1.0]{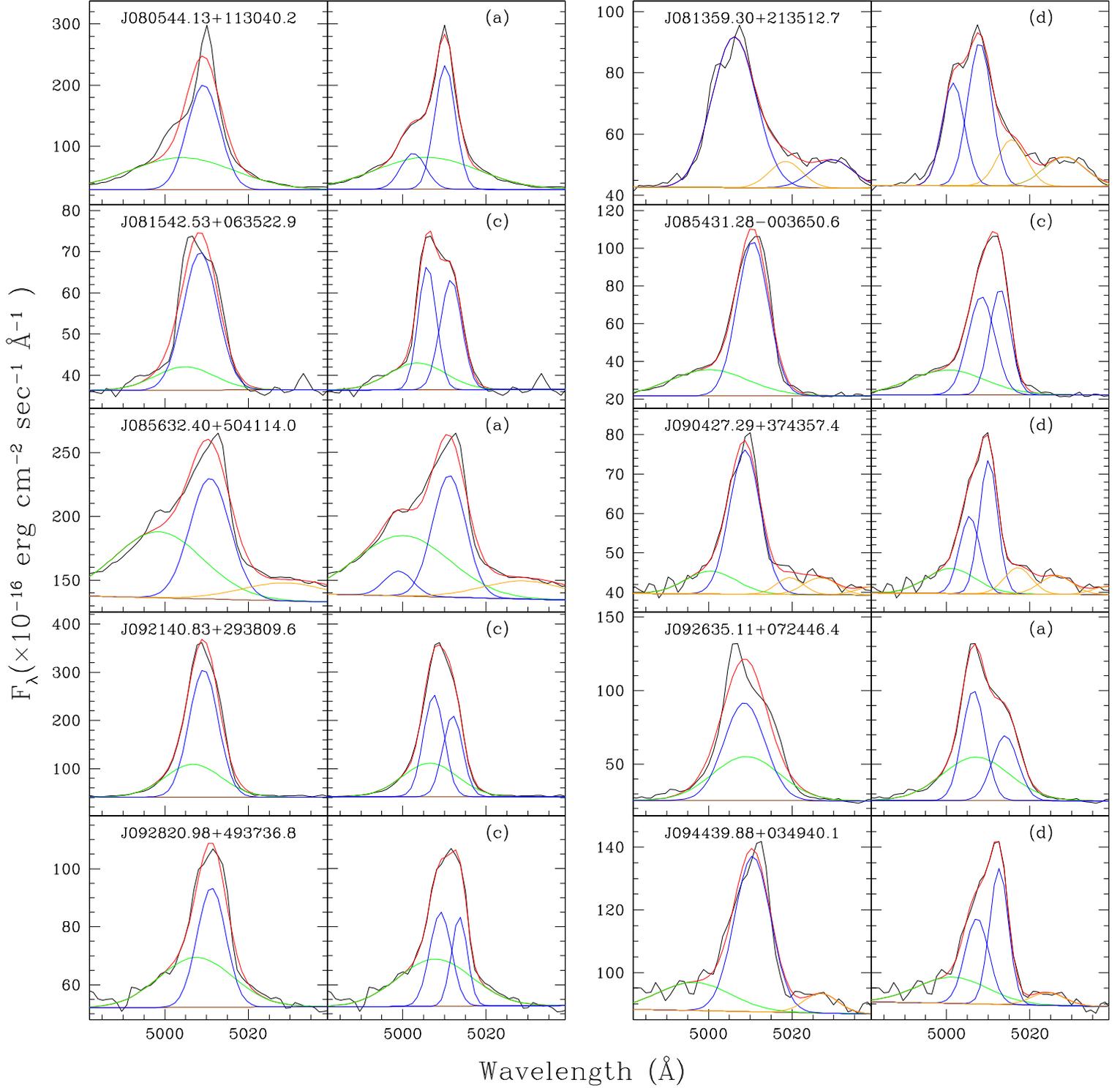}}
	\caption{Spectral decomposition of DAGN candidates, where black, red, brown, blue, green, and orange represent data, model, continuum, narrow line component, broad-line component, and Fe II lines, respectively. 
	For each object, left and right panel shows spectral decomposition with one [O III] plus broad-line components and two [O III] plus broad-line components, respectively. Alphabet on the right panel represents closely-matched representative spectral shape of the known DAGNs in Fig. 2.
}
\end{figure}

\begin{figure}[!hb]
\centerline{\includegraphics[scale=1.0]{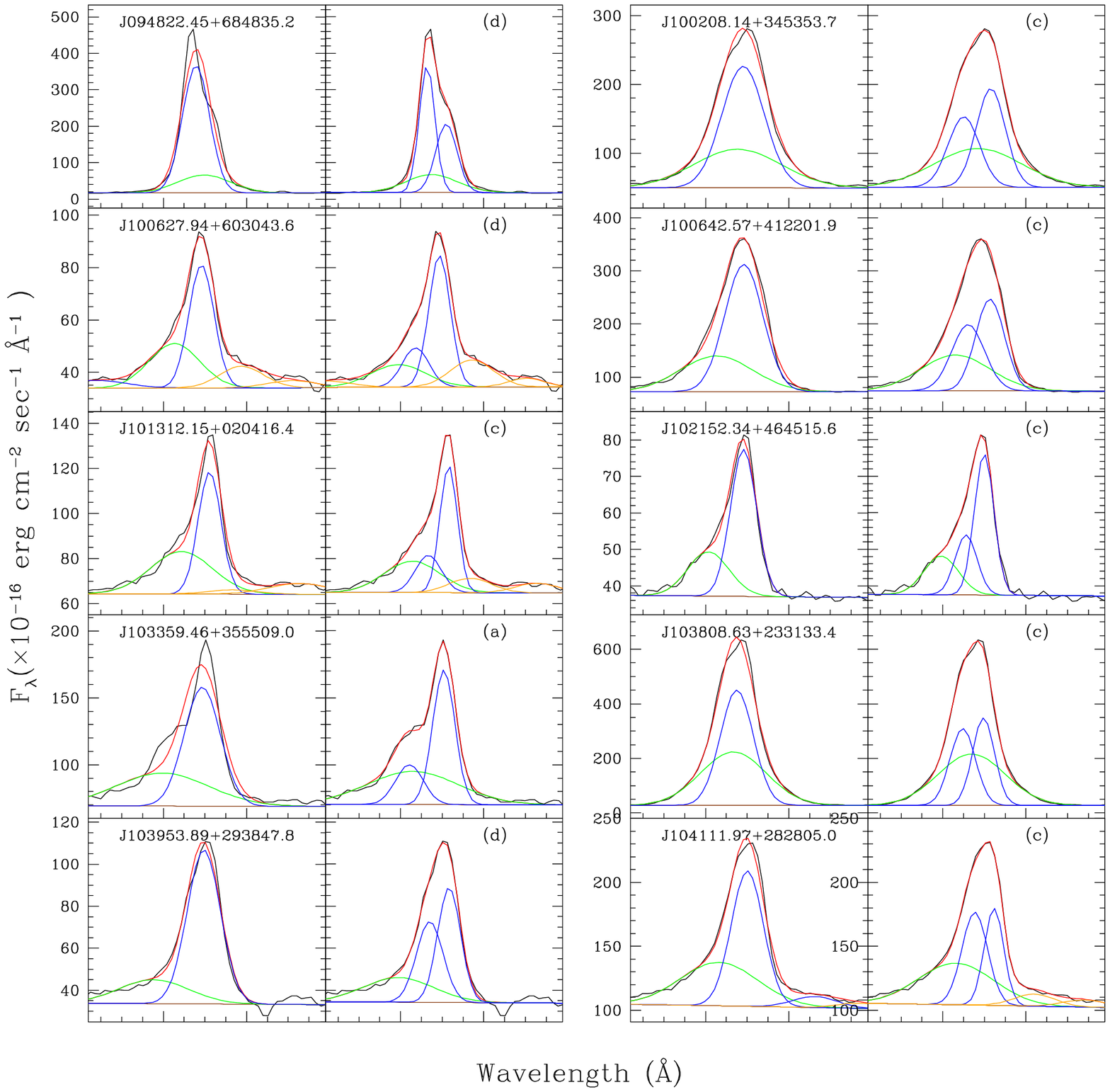}}
\end{figure}

\begin{figure}[!hb]
\centerline{\includegraphics[scale=1.0]{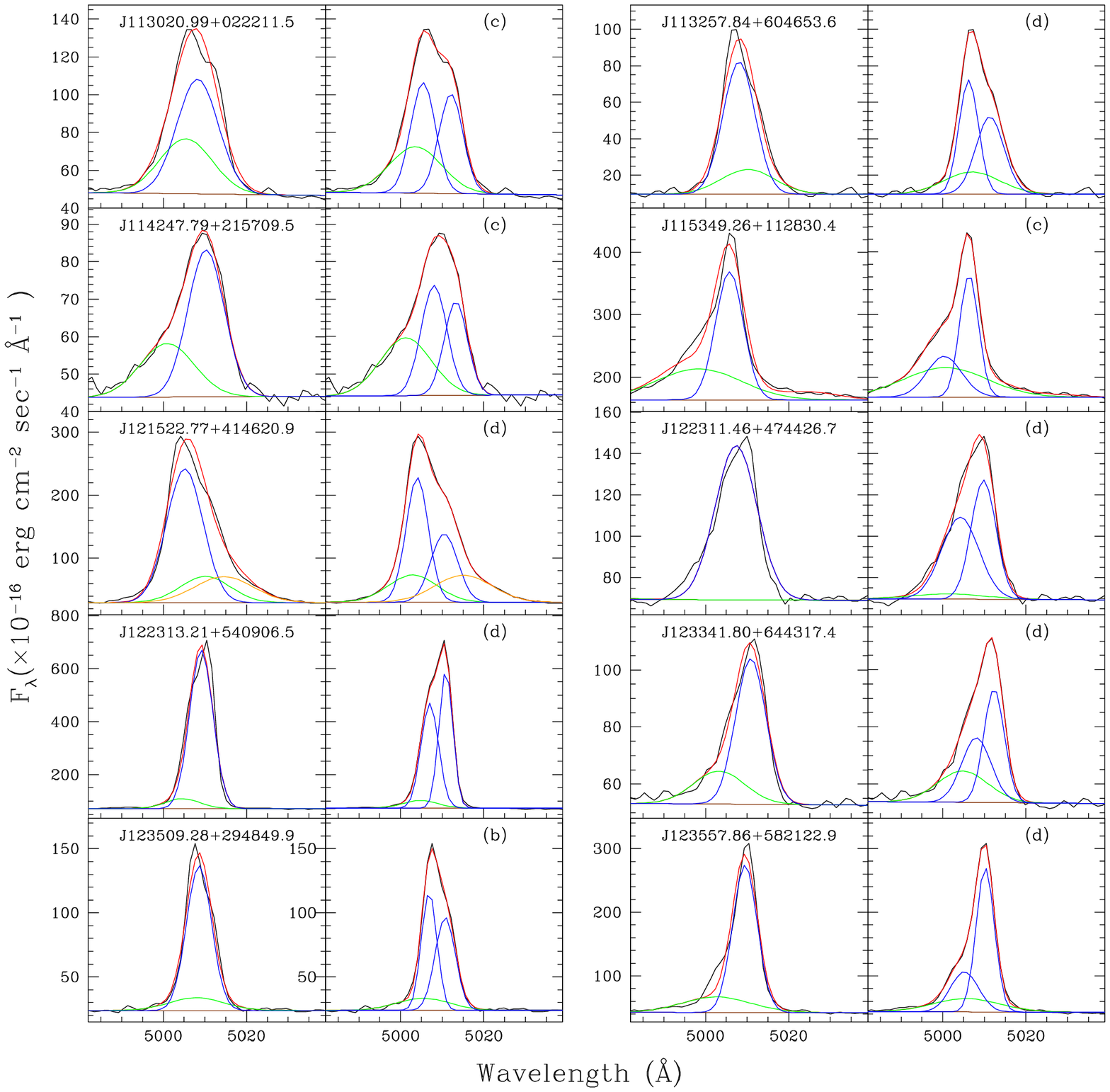}}
\end{figure}

\begin{figure}[!hb]
\centerline{\includegraphics[scale=1.0]{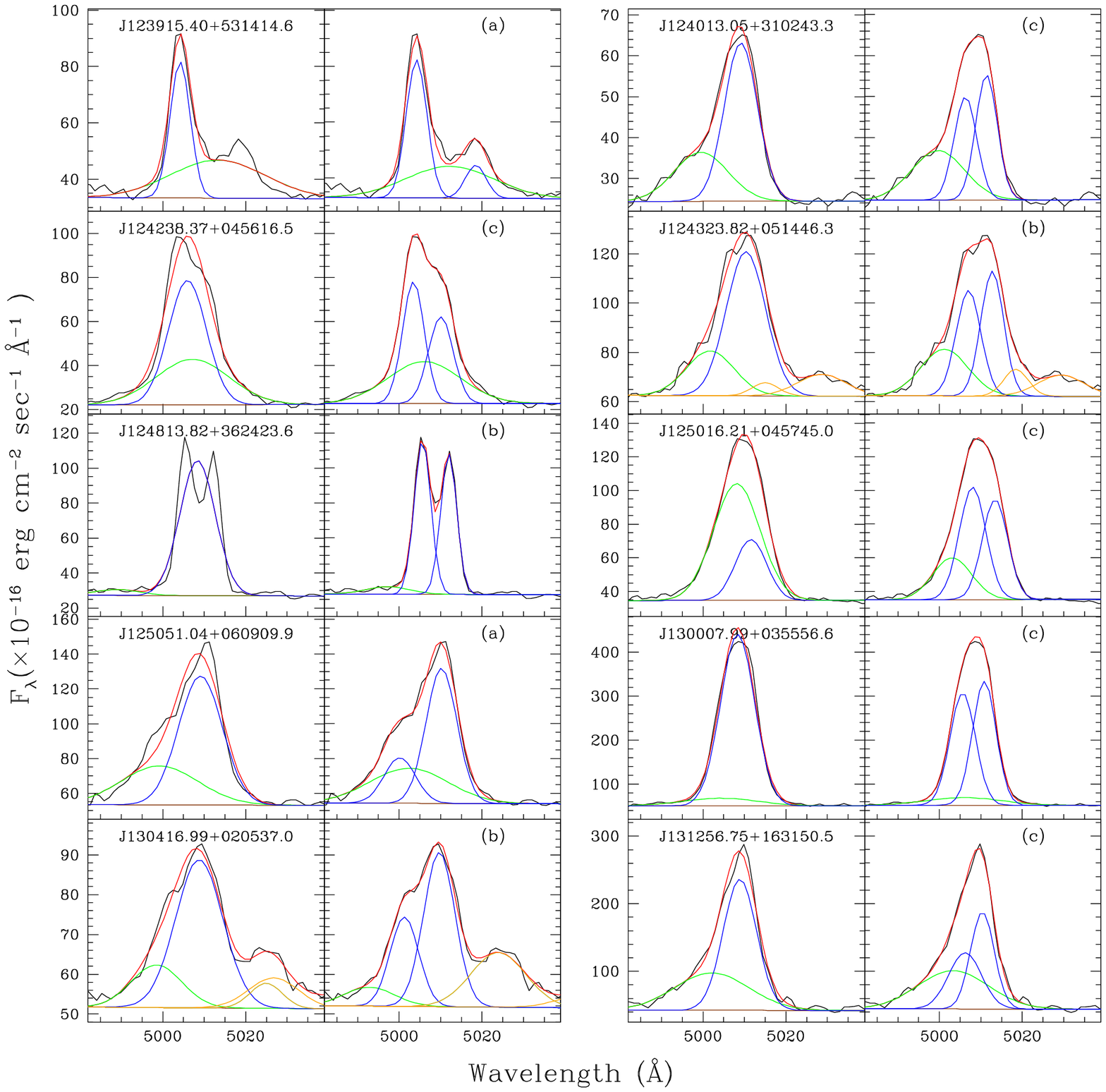}}
\end{figure}

\begin{figure}[!hb]
\centerline{\includegraphics[scale=1.0]{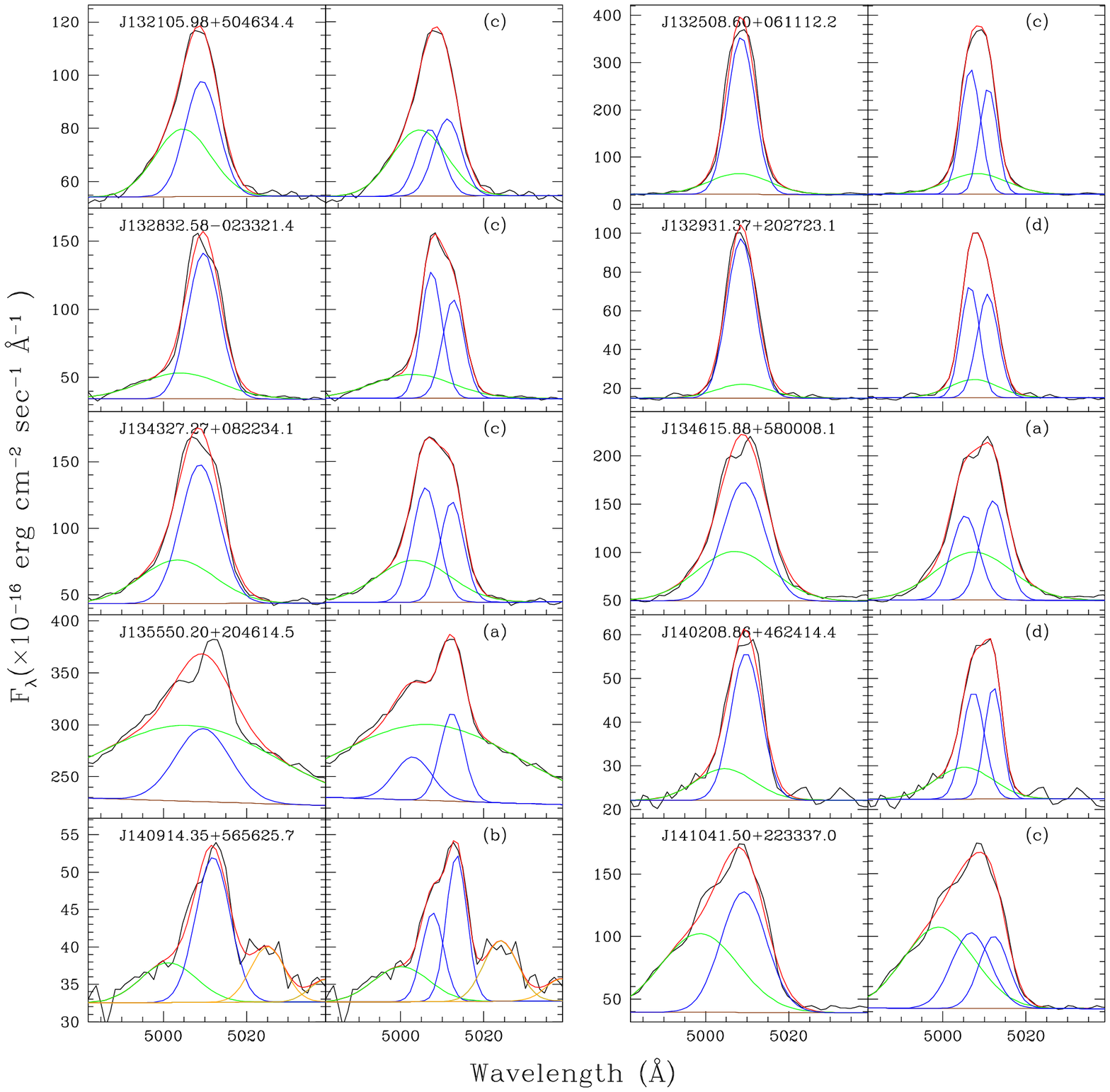}}
\end{figure}

\begin{figure}[!hb]
\centerline{\includegraphics[scale=1.0]{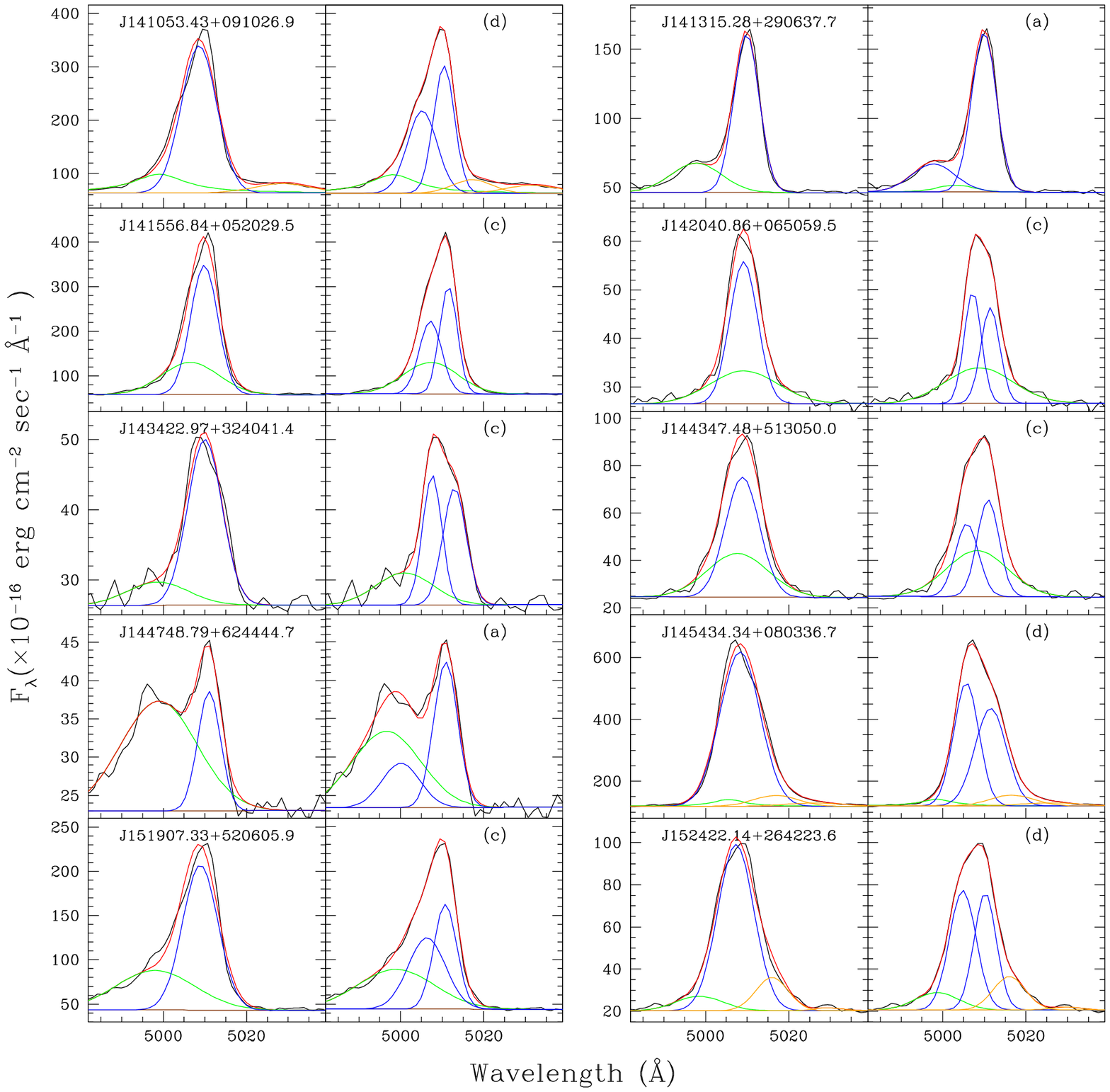}}
\end{figure}

\begin{figure}[!hb]
\centerline{\includegraphics[scale=1.0]{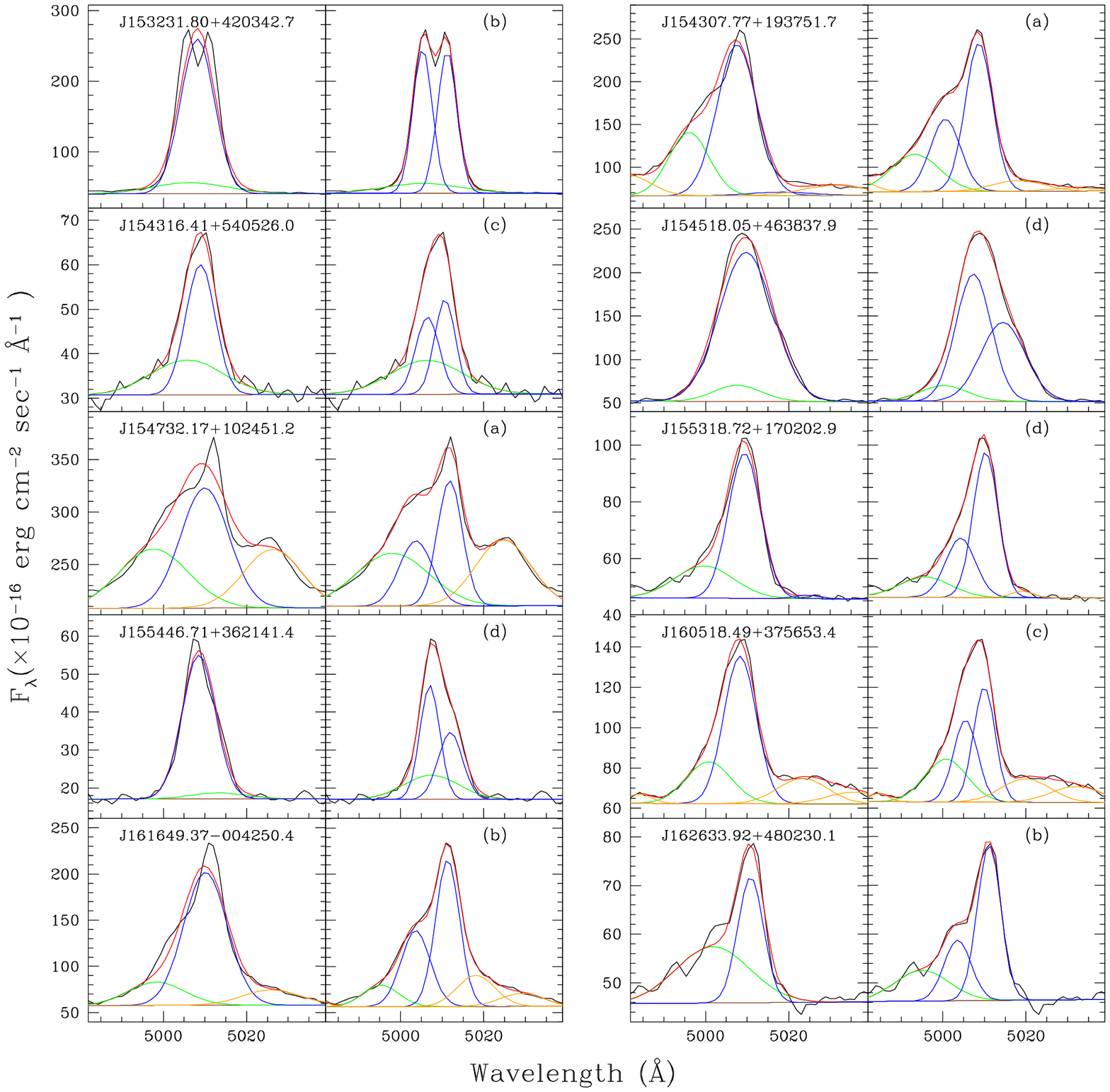}}
\end{figure}

\begin{figure}[!hb]
\centerline{\includegraphics[scale=1.0]{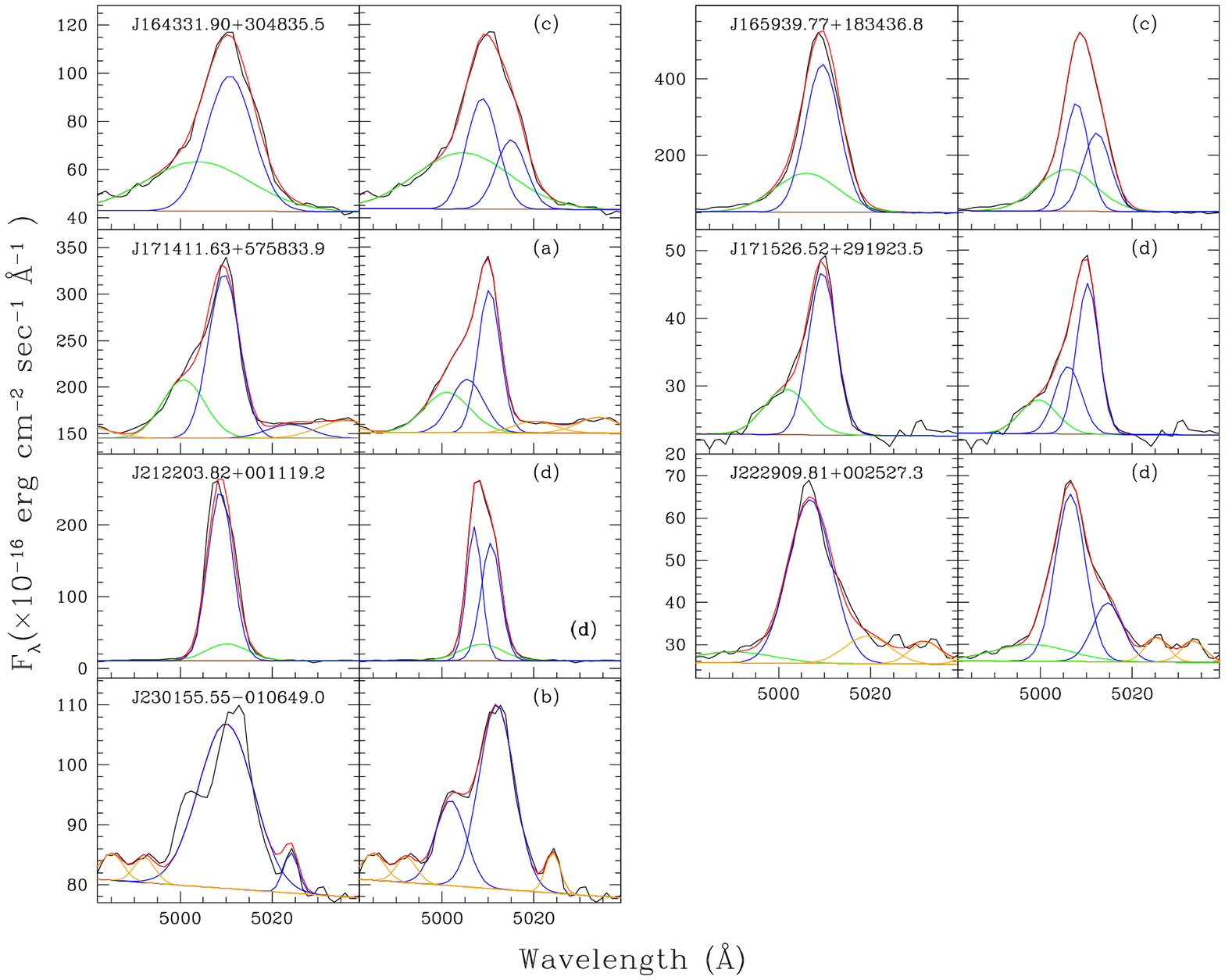}}
\end{figure}

\begin{figure}[!hb]
\centerline{\includegraphics[scale=1.0]{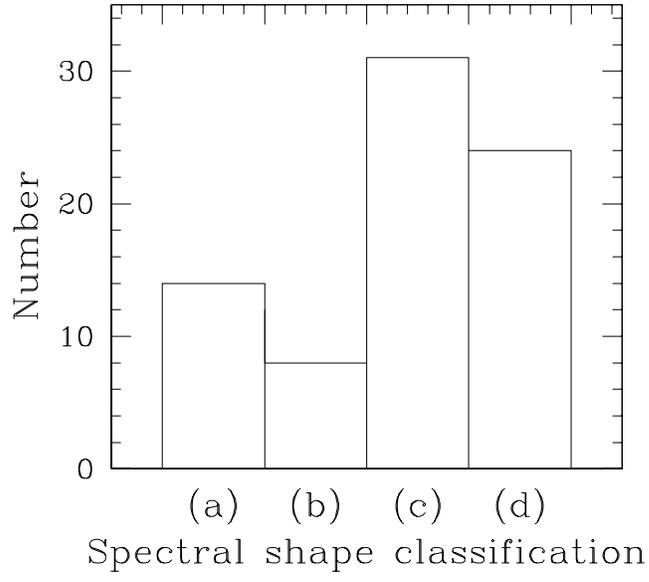}}
\caption{Histogram of spectral shape classification of the DAGN candidates
}
\end{figure}

\begin{figure}
\centerline{\includegraphics[scale=1.0]{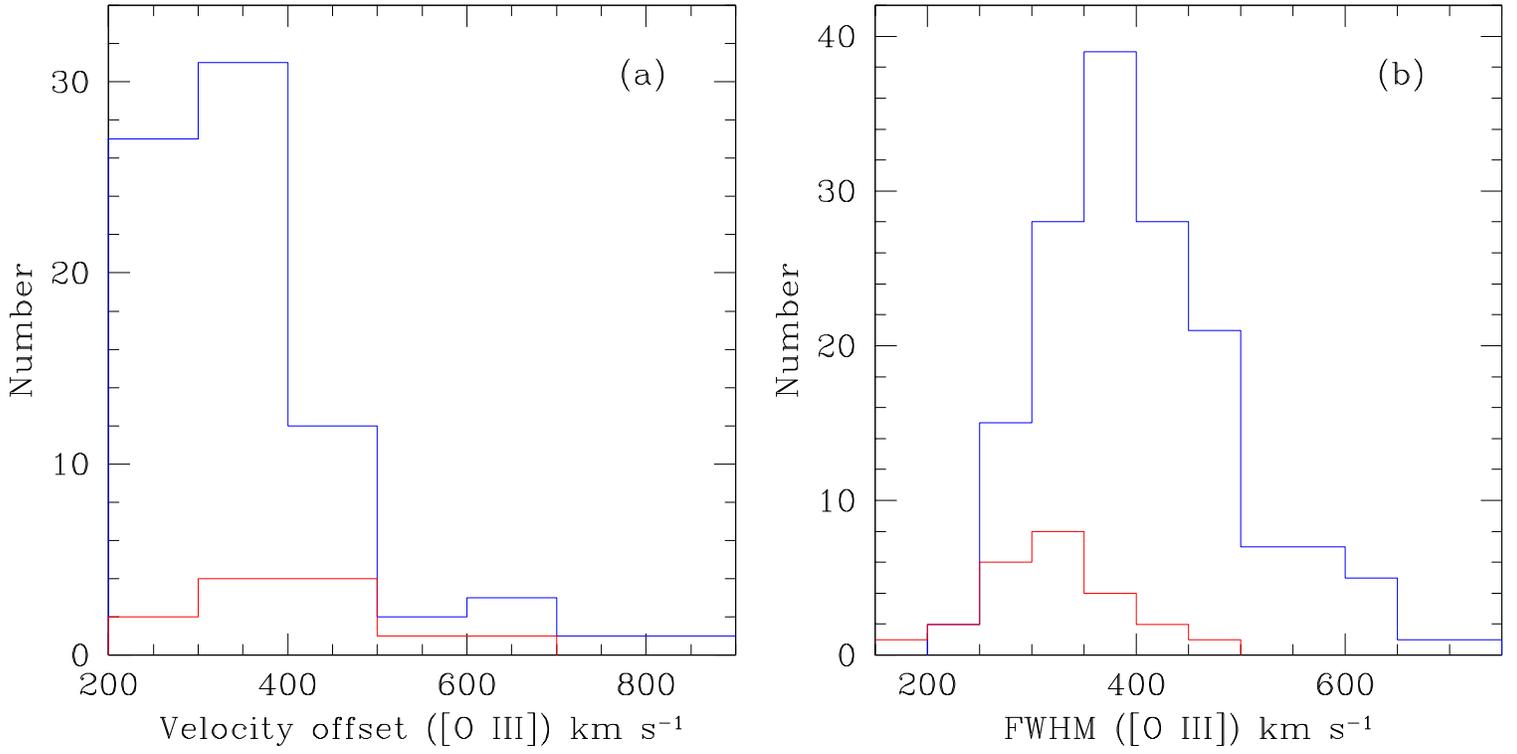}}
	\caption{Histograms of velocity offsets and line widths of the known DAGNs (red color) and DAGN candidates (blue color).
}
\end{figure}

\begin{figure}[!hb]
\centerline{\includegraphics[scale=0.98]{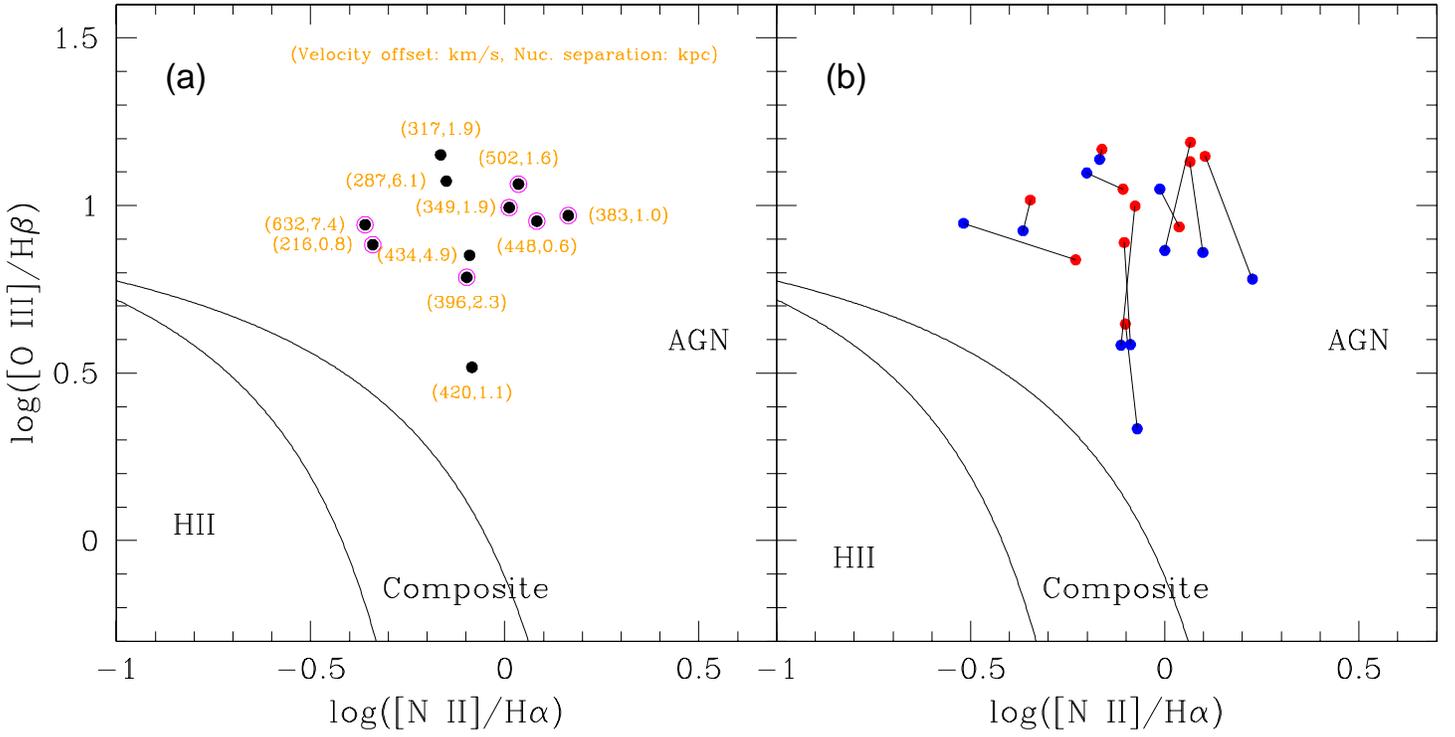}}
\caption{BPT-VO diagnostic diagram for known DAGNs. (a): integrated galaxy line ratios, and (b): line ratios for blueshifted component (blue) and redshifted component (red). Blueshifted and redshifted components are connected with a line.
}
\end{figure}

\begin{figure}[!hb]
\centerline{\includegraphics[scale=0.98]{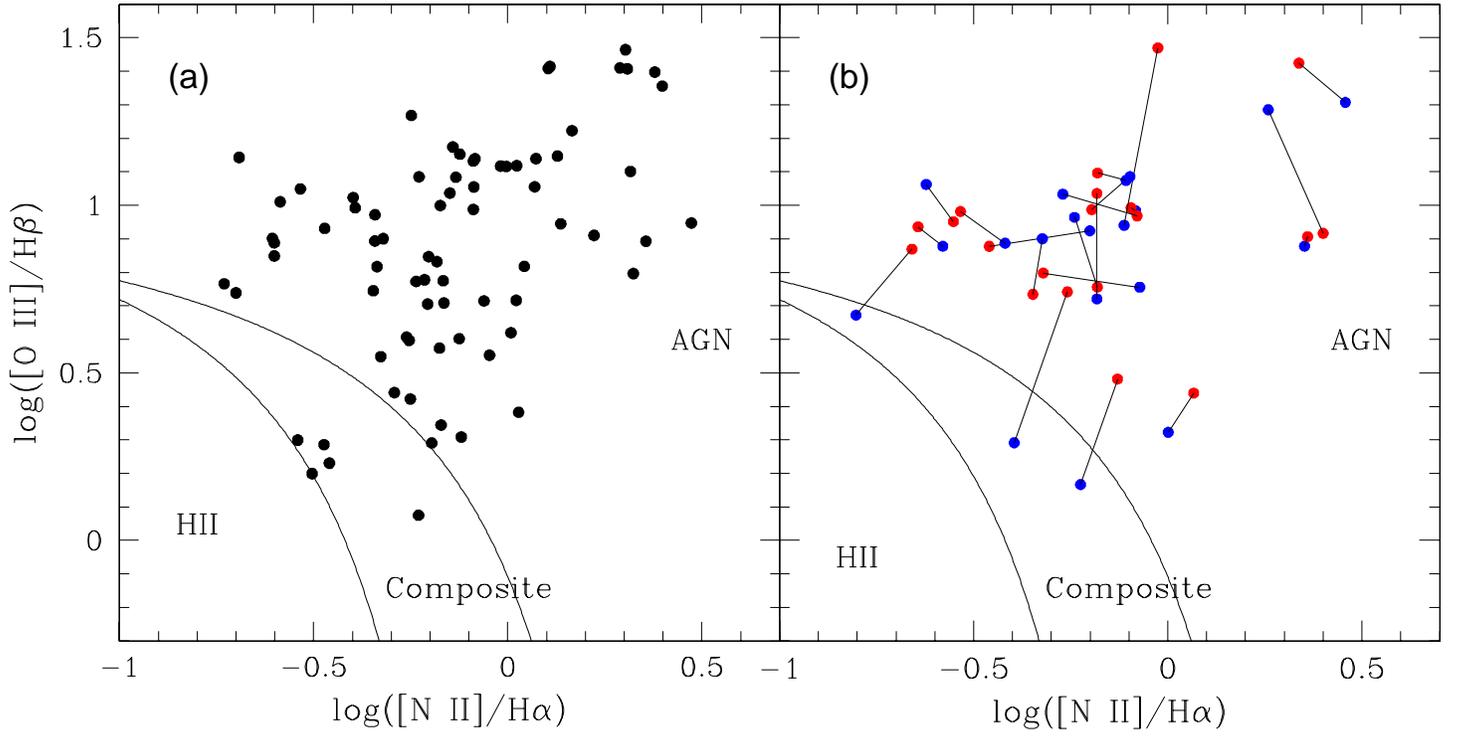}}
\caption{Same as Fig 6., but for the DAGN candidates.
}
\end{figure}

\begin{figure}[!hb]
\centerline{\includegraphics[scale=0.98]{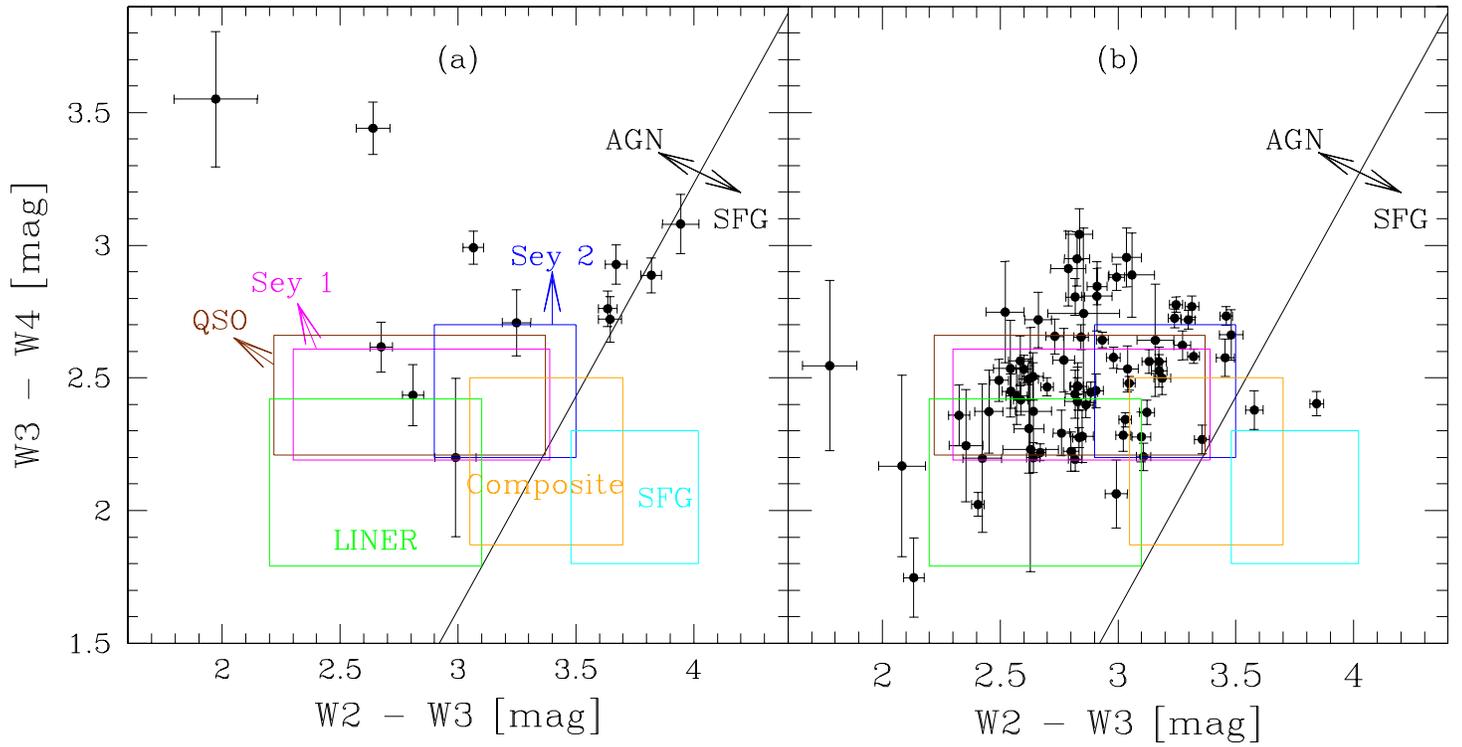}}
\caption{Mid-infrared diagnostic line diagram for known DAGNs (left panel) and for DAGN candidates (right panel). All but 2 DAGN candidates are powered by AGN.
}
\end{figure}

\begin{figure}[!hb]
\centerline{\includegraphics[scale=0.52]{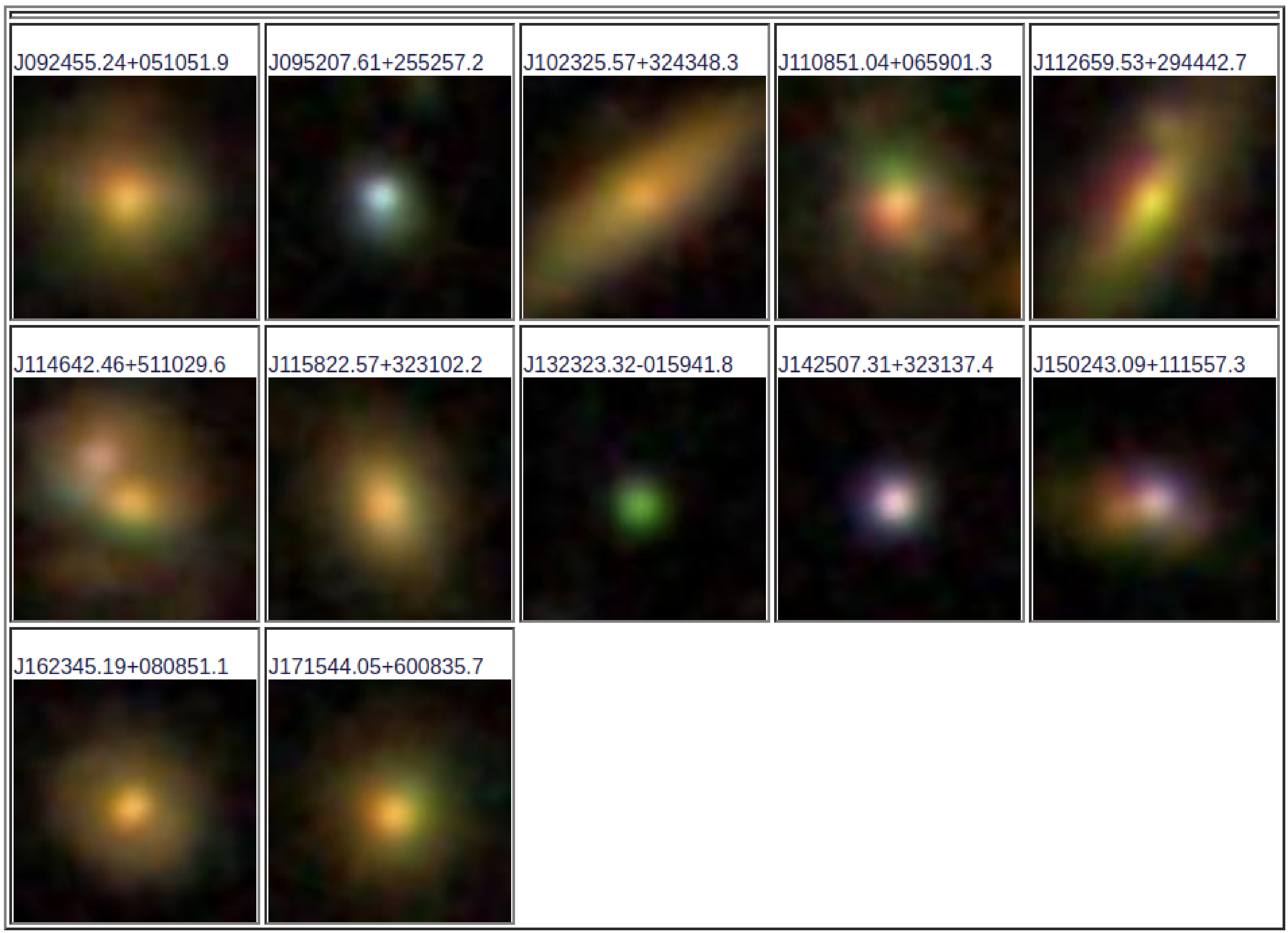}}
\caption{SDSS images of known DAGNs. FOV is 12\arcsec $\times$ 12\arcsec. North is top and east is to the left.
}
\end{figure}

\begin{figure}[!hb]
\centerline{\includegraphics[scale=1.2]{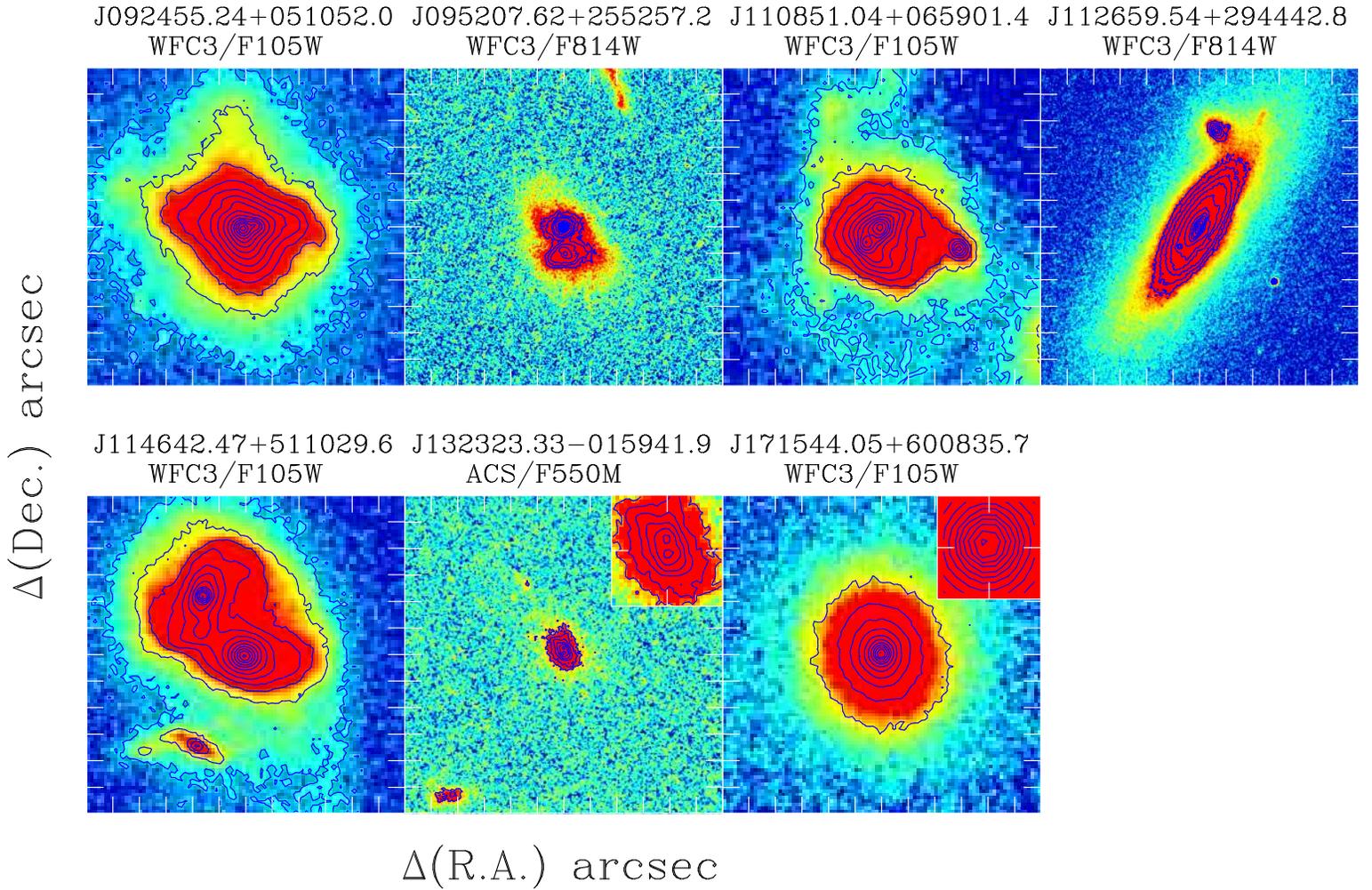}}
\caption{HST images of known DAGNs. HST camera and filter used are listed on top of each image below the object name. FOV is 12\arcsec $\times$ 12\arcsec. North is top and east is to the left.
}
\end{figure}

\begin{figure}[!hb]
\centerline{\includegraphics[scale=0.8]{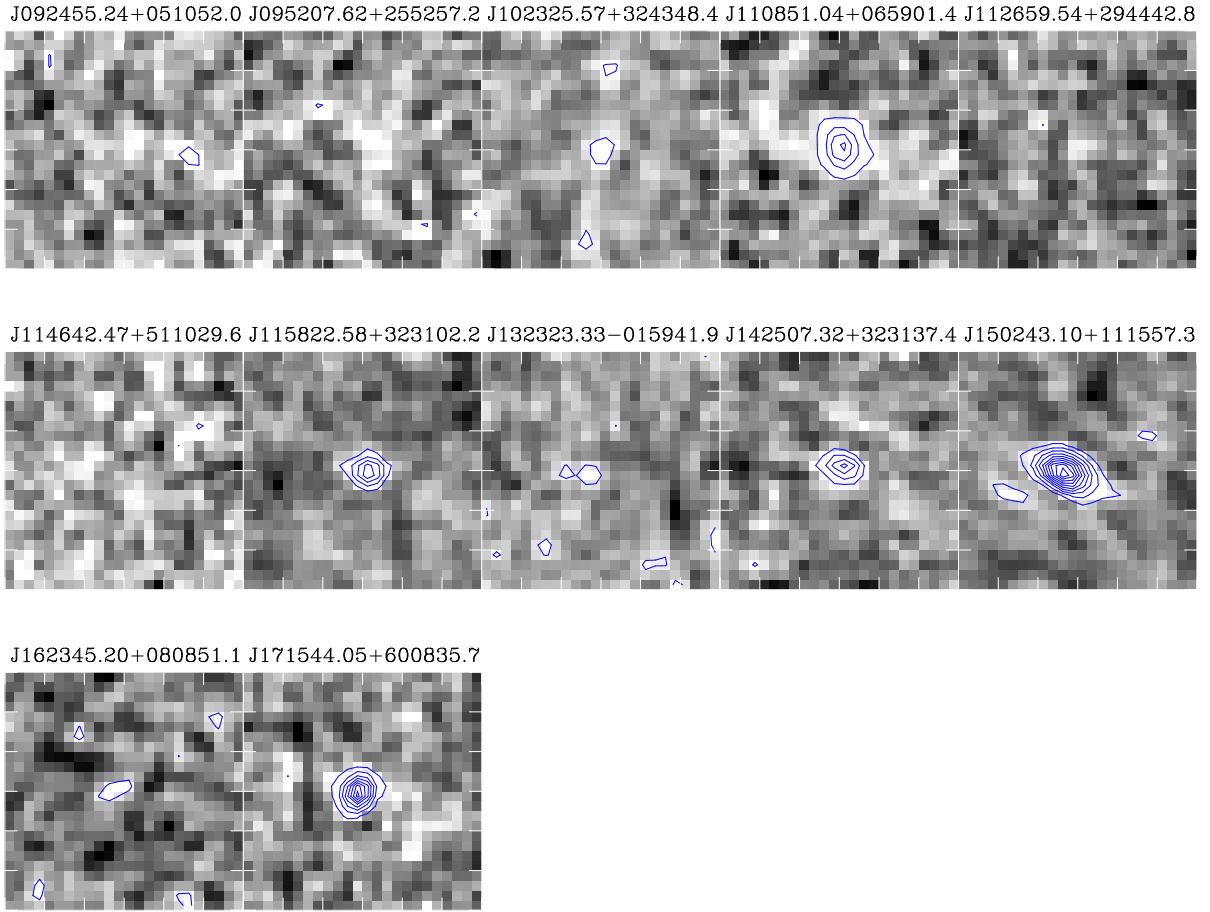}}
\caption{VLASS images of known DAGNs. FOV is 12\arcsec$\times$12\arcsec. Bottom contour level is 0.036 mJy and spaced multiples of it. North is top and east is to the left.
}
\end{figure}

\begin{figure}[!hb]
\centerline{\includegraphics[scale=0.52]{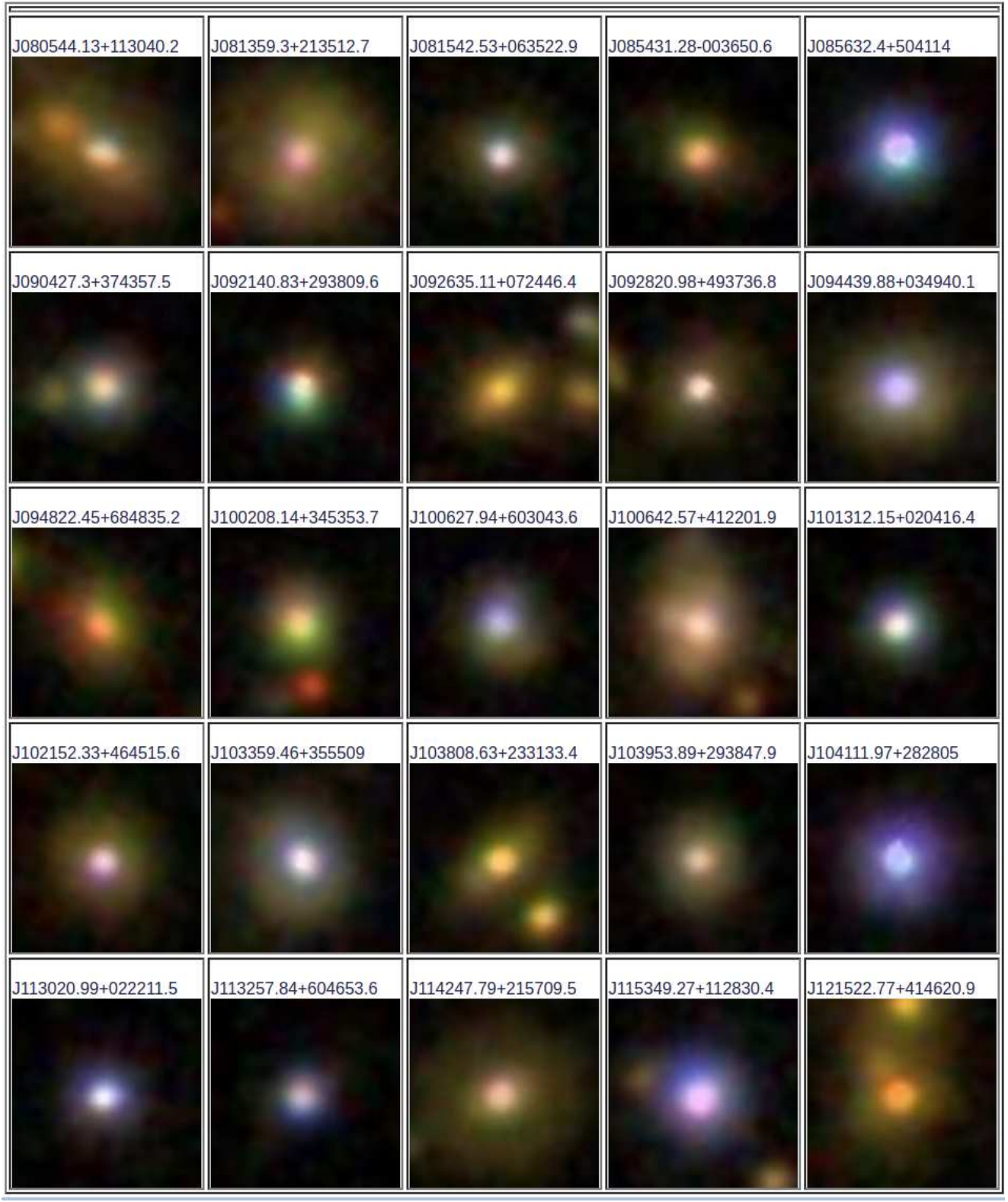}}
\caption{SDSS images of the DAGN candidates. FOV is 12\arcsec $\times$ 12\arcsec. North is top and east is to the left.
}
\end{figure}

\begin{figure}[!hb]
\centerline{\includegraphics[scale=0.52]{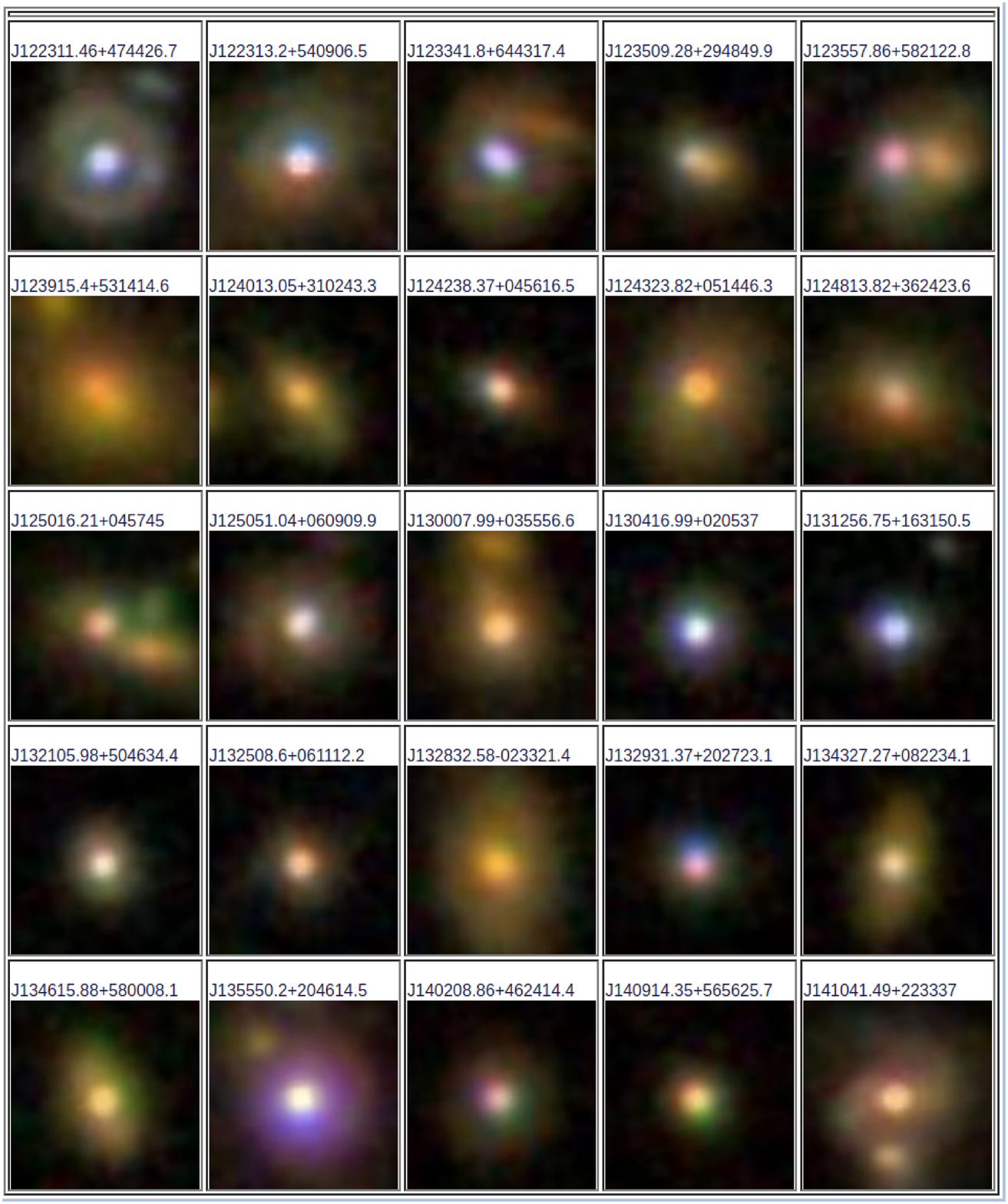}}
\end{figure}

\begin{figure}[!hb]
\centerline{\includegraphics[scale=0.52]{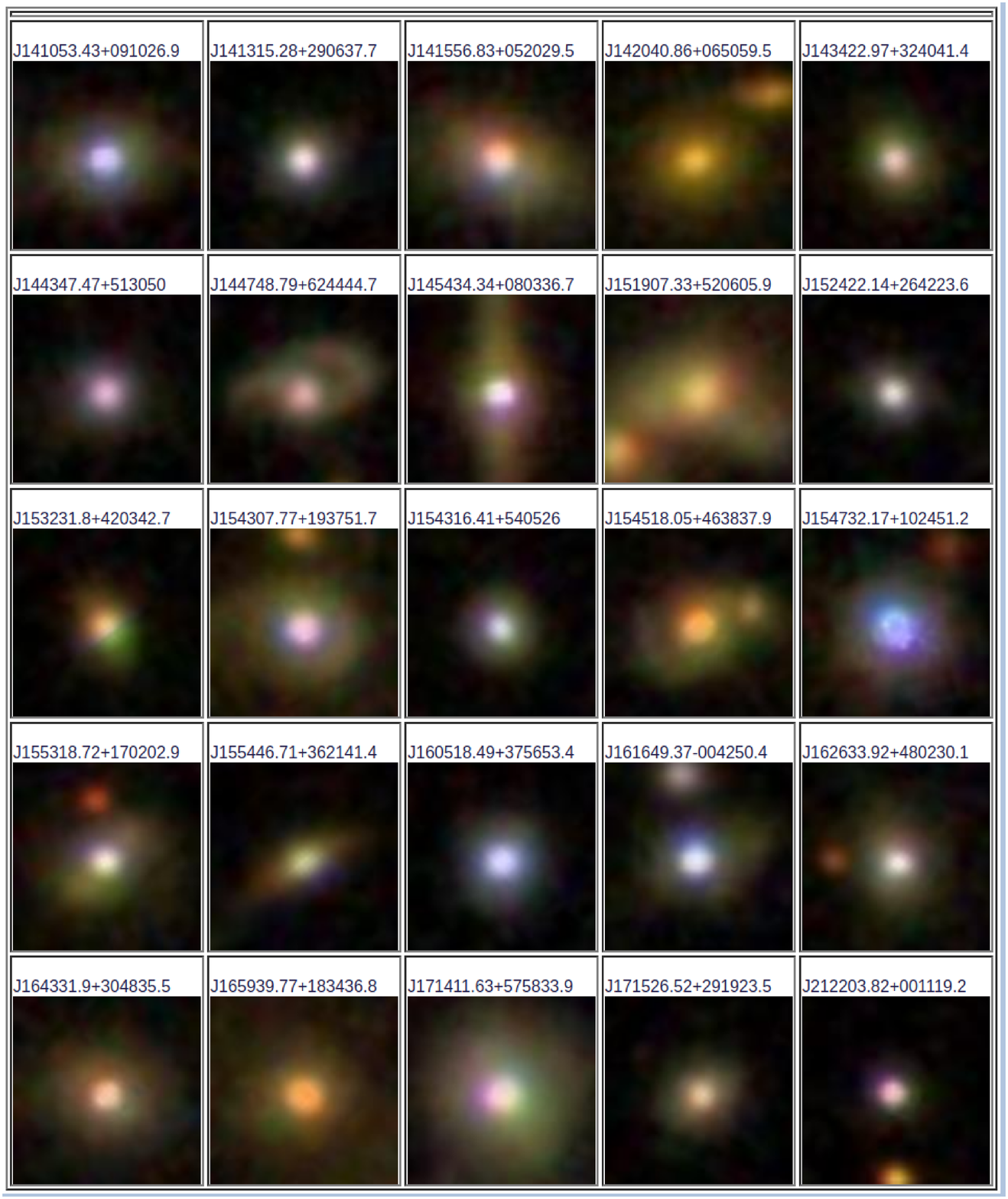}}
\end{figure}

\begin{figure}[!hb]
\centerline{\includegraphics[scale=0.52]{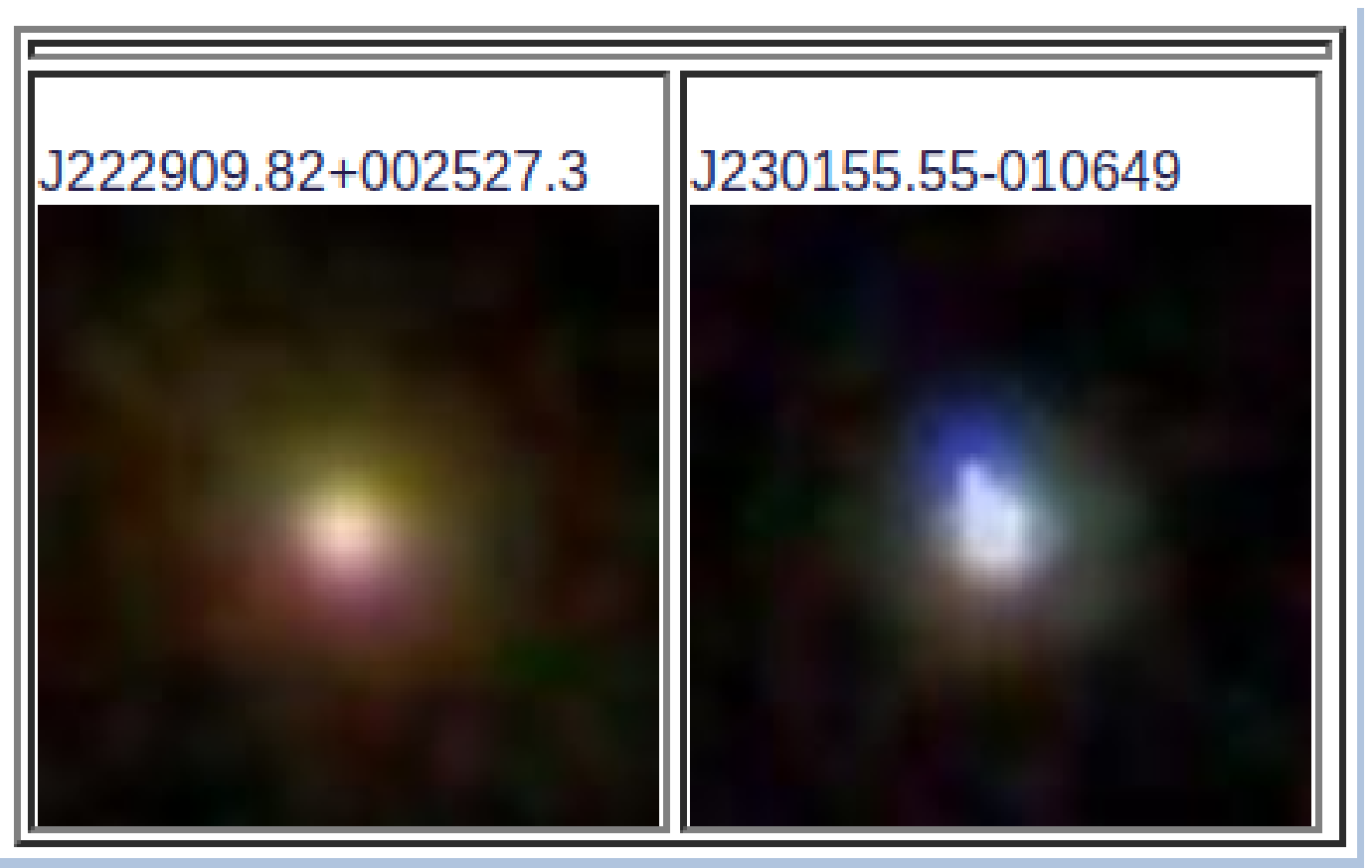}}
\end{figure}

\clearpage

\begin{figure}[!hb]
\centerline{\includegraphics[scale=1.2]{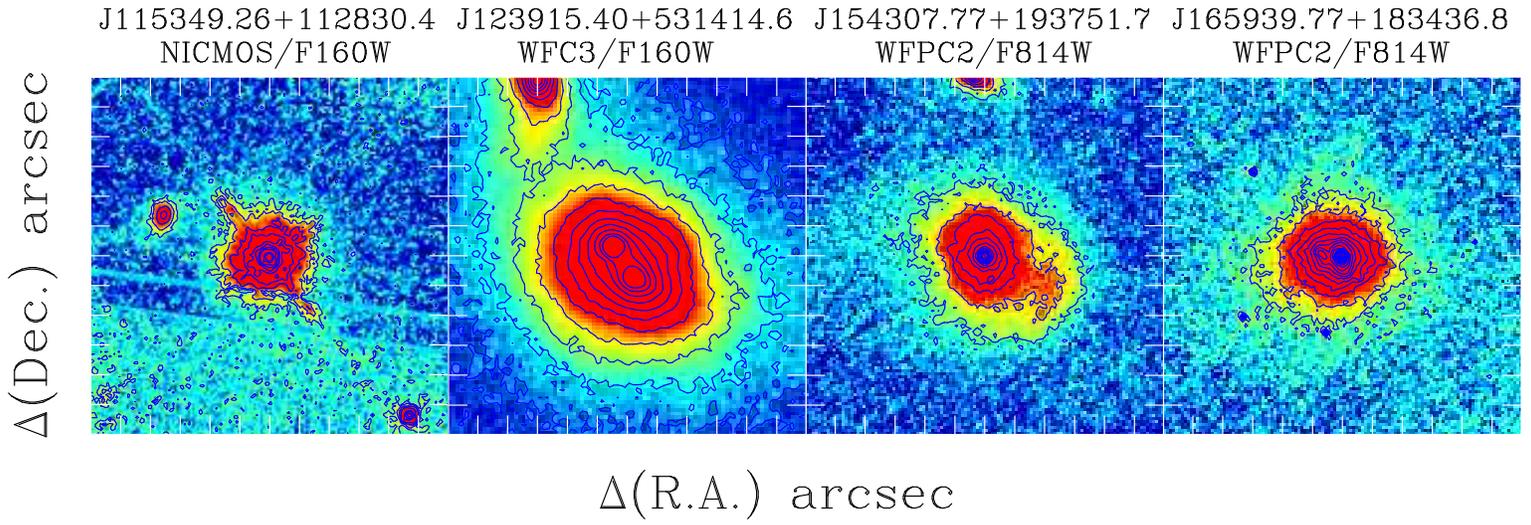}}
\caption{HST images of the DAGN candidates. HST camera and filter used are listed on top of each image below the object name. FOV is 12\arcsec$\times$12\arcsec. North is top and east is to the left.
}
\end{figure}

\begin{figure}[!hb]
\centerline{\includegraphics[scale=0.8]{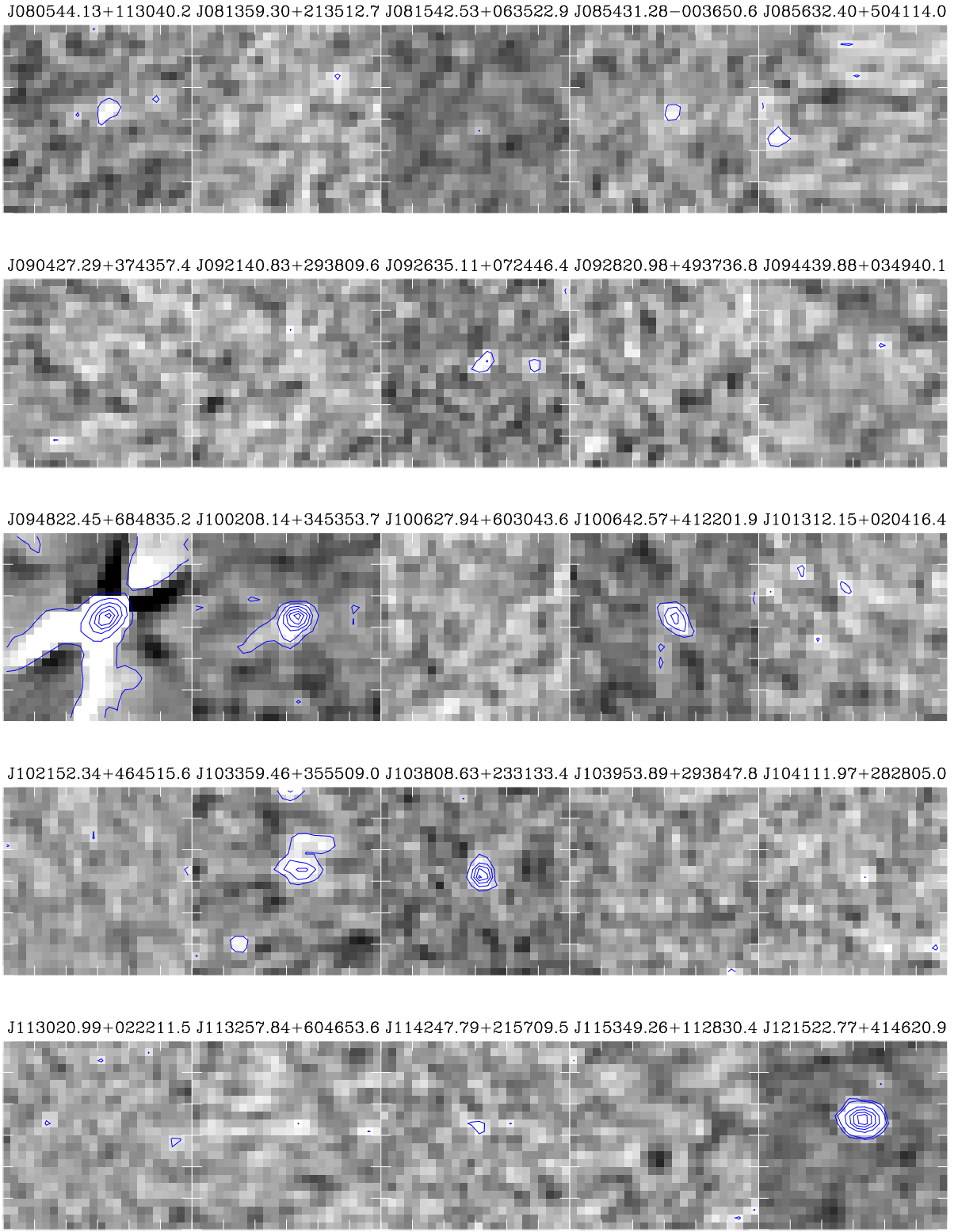}}
\caption{VLASS images of the DAGN candidates. FOV is 12\arcsec$\times$12\arcsec. Bottom contour level is 0.036 mJy and spaced multiples of it. North is top and east is to the left.
}
\end{figure}

\begin{figure}[!hb]
\centerline{\includegraphics[scale=0.8]{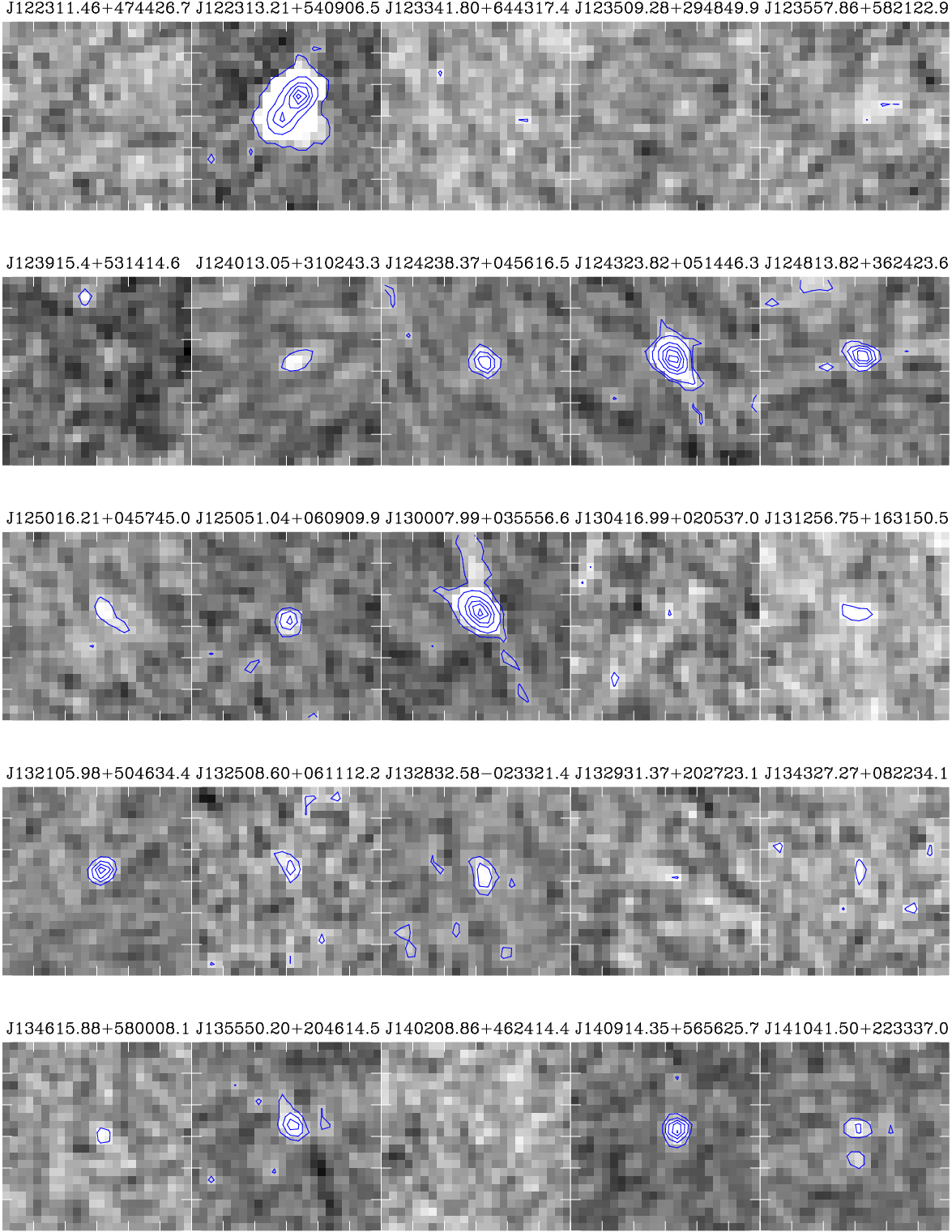}}
\end{figure}

\begin{figure}[!hb]
\centerline{\includegraphics[scale=0.8]{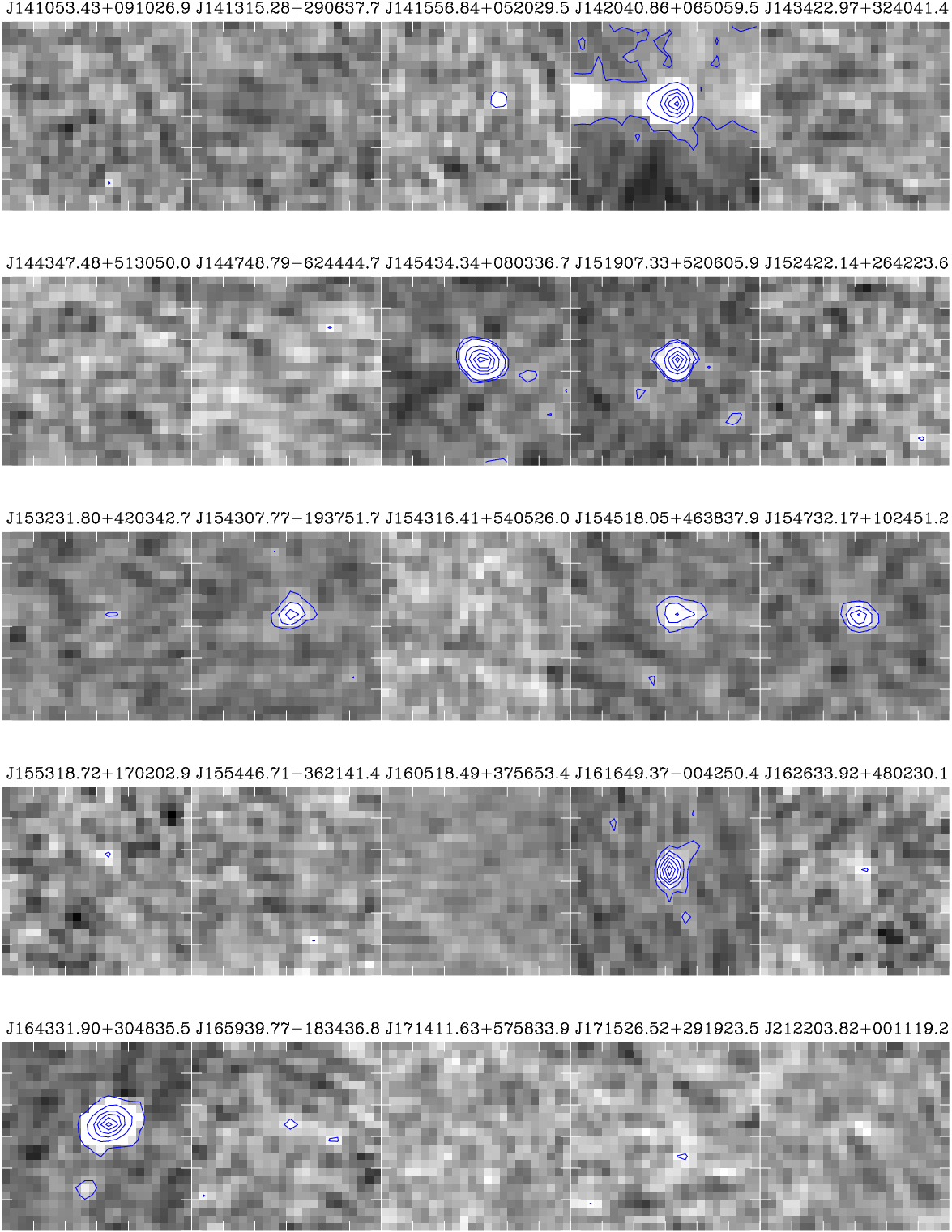}}
\end{figure}

\begin{figure}[!hb]
\centerline{\includegraphics[scale=0.8]{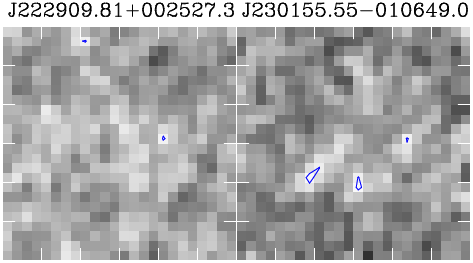}}
\end{figure}

\clearpage

\end{document}